\documentclass[prd,preprint,nofootinbib]{revtex4}
\usepackage{amssymb}
\usepackage{graphicx}
\def\Journal#1#2#3#4{{#1} {\bf #2}, #3 (#4)}

\def\NPB{{\em Nucl. Phys.} {\bf B}}
\def\PLB{{\em Phys. Lett.} {\bf B}}
\def\PRL{\em Phys. Rev. Lett.}
\def\PRD{{\em Phys. Rev.} {\bf D}}

\def\PR{\em Phys. Rep.}

\def\APP{\em Astropart. Phys.}
\def\APJ{\em Astrophys. J.}
\def\APJS{\em Astrophys. J. Suppl.}

\def\JHEP{{\em J. High Energy Phys.}}
\def\IJMP{{\em Int. J. Mod. Phys.}}
\def\JCAP{{\em J. Cosmo. Astropart. Phys.}}
\newcommand{\gev}{\,{\rm GeV}}
\newcommand{\tev}{\,{\rm TeV}}
\begin{document}
\title{ Suppression of the neutralino relic density with supersymmetric CP violation } 
\author{Takeshi Nihei}
  \email{nihei@phys.cst.nihon-u.ac.jp}
  \affiliation{ 
    Department of Physics, College of Science and Technology, Nihon University, 
    1-8-14, Kanda-Surugadai, Chiyoda-ku, Tokyo, 101-8308, Japan 
   }


\begin{abstract}
We study pair annihilations of the neutralino dark matter 
in the minimal supersymmetric standard model with CP violation. 
We consider the case that the higgsino mass and the trilinear scalar
couplings have CP-violating phases of order unity, 
taking a scenario that the scalar fermions in the first two generations 
are much heavier than those in the third generation to avoid a severe
constraint from experimental limits on electric dipole moments. 
It is found that, when the lightest neutralino ($\chi$) is bino-like, 
the cross sections of the $W$-boson pair production $\chi\chi$ $\to$ 
$W^+W^-$ and the lightest Higgs boson pair production $\chi\chi$ $\to$ 
$H^0_2 H^0_2$ for nonrelativistic neutralinos 
can be significantly enhanced by the phase of the higgsino mass. 
The relic density of the neutralino can be considerably suppressed 
by this effect. 
However, even this suppression is not enough to 
make bino-like dark matter consistent with a cosmological constraint. 
We also discuss the effect of CP violation on the positron flux 
from neutralino pair annihilations in the galactic halo. 
\end{abstract}
\preprint{January 2006}
\vspace{1cm}
\maketitle 
%
%
\section{Introduction}
%

Existence of a considerable amount of cold dark matter (CDM) 
in the present Universe has been regarded as a robust ingredient 
in recent astrophysics and cosmology \cite{Kolb-Turner}. 
In particular, 
the Wilkinson Microwave Anisotropy Probe (WMAP) has determined 
the relic abundance of CDM as \cite{WMAP} 
\begin{eqnarray}
\Omega_{\rm CDM}h^2 & = & 0.1126^{+0.008}_{-0.009}, 
\label{eqn:WMAP}
\end{eqnarray}
where $\Omega_{\rm CDM}$ is the CDM energy density normalized by
the critical density and 
$h$ $\approx$ 0.7 is the Hubble constant in units of 
100 km/sec/Mpc \cite{Hubble}. 
The relic density of the CDM is expected to be determined with more
accuracy by analyses of increasing WMAP data and 
forthcoming data from the future project Planck \cite{PLANCK}.

Identification of the CDM as a particle still remains to be settled. 
Among possible candidates, the lightest superparticle (LSP) 
in supersymmetric models is one of the most attractive candidates 
\cite{Jungman-etal,dm-recent-review}. 
In supersymmetric models, conservation of a discrete symmetry called R-parity 
is nessesary to avoid rapid proton decay, but it in turn
plays a crucial role to guarantee the stability of the LSP. 
In the minimal supersymmetric standard model (MSSM) \cite{MSSM}, 
the LSP is typically the lightest neutralino \cite{neutralino-dm} 
which is a linear combination of neutral gauginos and higgsinos
\begin{eqnarray}
\chi & = & \chi^0_1 \ = \ 
N_{11} \tilde{B} + N_{12} \tilde{W}^3 
+ N_{13} \tilde{H}^0_1 + N_{14} \tilde{H}^0_2, 
\label{eqn:chi-1}
\end{eqnarray}
where $\tilde{B}$ is the $U(1)_Y$ gaugino (bino),  
$\tilde{W}^3$ is the neutral $SU(2)_L$ gaugino (wino), and
$\tilde{H}^0_1$ and $\tilde{H}^0_2$ are the two neutral higgsinos
with opposite hypercharges. 
The coefficients $N_{1i}$ ($i$ $=$ $1,2,3,4$) are the elements of 
the 4$\times$4 unitary matrix $N$ which diagonalizes the neutralino 
mass matrix \cite{MSSM}.

A number of theoretical analyses 
on the relic density of the neutralino CDM in the MSSM 
have already been done extensively 
\cite{SWO,Eli-Ros-Lal,Drees-Nojiri,rd-ww,Gondolo-Gelmini,Lopez-Nano-Yuan,Baer-Brhlik,NRR1,RRN,NRR2,oh2-recent}. 
However, most analyses have been done with an assumption of 
no CP violation in supersymmetric parameters. 
CP violation is not only observed in particle physics
but also expected to play an essential role 
to explain baryon asymmetry in the Universe. 
Since it is known that the standard source of CP violation in the
Cabibbo--Kobayashi--Maskawa mixing matrix is not enough to produce 
required asymmetry of order $10^{-10}$, 
supersymmetric CP violation is expected 
to gerenate the correct amount of baryon asymmetry 
\cite{electroweak-baryogenesis,Affleck-Dine-baryogenesis}. 
Thus, it is generically important to include the effect of 
supersymmetric CP violation. 
When the supersymmetric CP phases are of order unity, 
they typically lead to too large electric dipole moments of
the neutron, the electron and ${}^{199}$Hg atom 
to satisfy experimental limits \cite{dn-limit,de-limit,Hg-limit}
if masses of superparticles are within a TeV scale. 
However, if superparticles in the first two generations are much heavier 
than 1 TeV, the supersymmetric CP phases of order unity are still allowed.

On the other hand, 
a complete analysis of the neutralino relic density $\Omega_{\chi} h^2$ 
including all the relevant effects 
in the MSSM with CP violation is still missing. 
The effect of CP violation on 
the neutralino pair annihilation cross section into the fermion pairs 
through the sfermion exchange was studied in Ref.~\cite{FOS}. 
In Ref.~\cite{GF}, the effects of a CP-violaing phase in the trilinear 
scalar couplings on the neutralino pair annihilation cross section 
were examined, including all the final states and 
taking into account the scalar and pseudoscalar mixing 
in the neutral Higgs sector \cite{Pilaftsis,S_PS_mixing_2}. 
Recently, the dependence of $\Omega_{\chi} h^2$ on CP-violating 
phases through supersymmetric loop corrections to the bottom-quark 
mass was examined in Ref.~\cite{GINS}. 
Partial wave treatment in the presence of CP violation was 
studied in Ref.~\cite{ALNS}. 
An analysis related with electroweak baryogenesis was done 
in Ref.~\cite{BCMMW}.

In this paper, we extend our previous analysis \cite{nihei-sasagawa}
and perform a reanalysis on the effects of CP-violating 
complex phases in the MSSM on pair annihilation cross sections of 
the lightest neutralino and its relic density in the present Universe. 
We consider the case that the higgsino mass and the trilinear scalar
couplings have CP-violating phases of order unity, 
taking a scenario that the scalar fermions in the first two generations 
are much heavier than those in the third generation to avoid a severe
constraint from experimental limits on electric dipole moments. 
We include all the contributions to the neutralino annihilation cross 
section at the tree level, taking into account the scalar and pseudoscalar
mixing in the neutral Higgs sector. 
In the absence of CP violation, it is known that 
fermion pair productions $\chi\chi$ $\to$ $f\bar{f}$ 
usually give the largest contribution to the total cross section 
for a bino-like LSP. 
In the present paper,
we shall show that, unlike the case without CP violation, 
the $W$-boson pair production $\chi\chi$ $\to$ $W^+W^-$ 
or the lightest Higgs boson pair production $\chi\chi$ $\to$ $H^0_2 H^0_2$ 
can give the largest contribution in the presence of CP violation. 
We also discuss the effect of CP violation on the positron flux 
from neutralino pair annihilations in the galactic halo.

This paper is organized as follows. 
In section \ref{sec:mssm}, we describe the structure of 
the MSSM with supersymmetric CP violation, and present 
the relevant interactions. 
In section \ref{sec:cross-section}, 
we discuss the effect of CP violation on 
the cross section of the lightest neutralino pair annihilation.
In section \ref{sec:relic-density}, 
we briefly review the formalism to compute the relic density
of the neutralino. 
In section \ref{sec:positron}, we summarize the procedure to obtain
the positron flux. 
In section \ref{sec:numerical-results}, we present our numerical results. 
Finally concluding remarks are given in Section \ref{sec:conclusions}.

%
%
\section{The MSSM with supersymmetric CP violation}
\label{sec:mssm}
%
In the present work, we consider a general MSSM \cite{MSSM} 
in which all the interactions are specified 
by the following input parameters at the weak scale 
\begin{eqnarray}
M_1, \ M_2, \ M_3, \ \mu, \ m_0, \ A,
\ \tan \beta, \ m_A, 
\label{eqn:mssm-parameters}
\end{eqnarray}
where $M_1$, $M_2$ and $M_3$ are the mass parameters for 
the bino, the wino and the gluino, respectively. 
The parameter $\mu$ represents the Higgs mixing mass, and 
$m_0$ is the supersymmetry breaking common mass parameter 
for the sfermions in the third generation. 
The corresponding mass parameters for the first two generations
are assumed to be 10 TeV to suppress the electric dipole moments
\footnote{The sfermion mass spectrum considered here naturally arises
in a minimal supersymmetric $SO(10)$ GUT model \cite{SO10-raby}. }. 
The parameter $A$ is the common trilinear scalar coupling for the third 
generation, while the ones for the first two generations are neglected. 
The ratio of the vacuum expectation values of the two neutral Higgs fields
is denoted by $\tan \beta$. 
The quantity $m_A$ is a parameter 
which coincides with the pseudoscalar Higgs mass 
in the absence of CP violation \cite{Pilaftsis,S_PS_mixing_2}. 
In eq.~(\ref{eqn:mssm-parameters}), 
$M_i$ ($i$ $=$ $1,2,3$), 
$\mu$ and $A$ can have CP-violating phases in general. 
For the gaugino masses, we assume the GUT relation 
\begin{eqnarray}
\frac{3}{5} \frac{M_1}{g^{\prime \,2}} \ = \ 
\frac{M_2}{g^2} \ = \ \frac{M_3}{g_s^2}, 
\label{eqn:gut-relation}
\end{eqnarray}
where $g'$, $g$ and $g_s$ are
the gauge coupling constants for $U(1)_Y$, $SU(2)_L$ and $SU(3)_C$ gauge 
groups, respectively. 
With this relation, the neutralino LSP can be bino-like 
($|M_1|$ $\ll$ $|\mu|$), higgsino-like ($|M_1|$ $\gg$ $|\mu|$)
or their mixture ($|M_1|$ $\approx$ $|\mu|$).

In the present analysis, we assume that the gaugino masses $M_i$ are
real, and examine the effects of the CP-violating phases of $\mu$ and $A$ 
\begin{eqnarray}
\mu & = & |\mu|\exp(i\theta_{\mu}), 
 \ \ A \ = \ |A| \exp(i\theta_{A}). 
\label{eqn:phase}
\end{eqnarray}
These phases induce imaginary parts in the mass matrices for 
the neutralinos, the charginos and the sfermions. 
They also induce mixings between scalar Higgs fields 
and a pseudoscalar Higgs field
through radiative corrections \cite{Pilaftsis,S_PS_mixing_2}. 
The MSSM contains two scalar neutral Higgses $\phi^0_1$ and $\phi^0_2$ 
with hypercharges $\frac{Y}{2}(\phi^0_1)$ $=$ $-\frac{Y}{2}(\phi^0_2)$ 
$=$ $-\frac{1}{2}$ and a pseudoscalar neutral Higgs $A^0$ as physical fields. 
In general, nonzero phases in $\mu$ or $A$ induce $\phi^0_1$--$A^0$ and 
$\phi^0_2$--$A^0$ mixings at the one-loop level, so that 
the 3$\times$3 mass-squared matrix ${\cal M}_H^2$ for the neutral Higgs
fields has to be diagonalized. 
We calculate the Higgs mass-squared matrix with the one-loop effective 
potential, including the contributions of the third generation 
fermions and sfermions. 
The mass eigenstates $H^0_r$ ($r$ $=$ 1, 2, 3) are related with the 
CP eigenstates ($\phi^0_1$,$\phi^0_2$,$A^0$) by a 3$\times$3 rotation 
matrix $O$ as follows
\begin{eqnarray}
\left( 
\begin{array}{c}
H^0_1 \\ H^0_2 \\ H^0_3 
\end{array}
\right)
 & = & 
O
\left( 
\begin{array}{c}
\phi^0_1 \\ \phi^0_2 \\ A^0
\end{array}
\right). 
\label{eqn:diag_higgs}
\end{eqnarray}
The Higgs boson mass eigenvalues are obtained as $O {\cal M}_H^2 O^T$
$=$ diag($m_{H^0_1}^2$,$m_{H^0_2}^2$,$m_{H^0_3}^2$). 
In eq.~(\ref{eqn:diag_higgs}), $H_2^0$ is defined as the lightest Higgs. 
The heavier Higgses $H_1^0$ and $H_3^0$ are defined such that 
$|O_{13}|$ $\leq$ $|O_{33}|$. 
In the case of $\theta_{\mu}$ $=$ $\theta_A$ $=0$, it follows that
$H_1^0$ ($H_2^0$) is the heavier (lighter) scalar Higgs field, and 
$H_3^0$ is the pseudoscalar Higgs field. 
Note that the parameter $m_A$ in eq.~(\ref{eqn:mssm-parameters})
is not a mass eigenvalue in general. 
The scalar--pseudoscalar mixing in the neutral Higgs sector implies 
$O_{r3}$ $\neq 0$ ($r=1,2$) so that interactions of the neutral Higgs
bosons are significantly modified \cite{Pilaftsis,S_PS_mixing_2}.

Our phase convention in the Higgs sector is defined as follows. In addition 
to $\theta_{\mu}$ and $\theta_A$, the coefficient of a Higgs mixing term 
($\phi_1^0 \phi_2^0$) can have a complex phase $\theta_{12}$ 
(See Ref. \cite{Pilaftsis}). 
In general these phases induce a nonvanishing relative phase $\xi$ between 
the vacuum expectation values of the two Higgs fields through the relevant 
tadpole minimum condition. However, $\theta_{12}$ and $\xi$ are not 
separately physical quantities, and only their sum $\xi+\theta_{12}$ is 
rephasing invariant. We adopt a convention that $\xi=0$ at the one-loop 
level so that nonvanishing $\theta_{12}$ will be induced by the tadpole 
condition in the presence of nonvanishing $\theta_{\mu}$ or $\theta_A$.

In the present work, we focus on the $W$-boson pair production
$\chi\chi$ $\to$ $W^+W^-$ and the lightest Higgs boson pair production 
$\chi\chi$ $\to$ $H^0_2 H^0_2$ . 
For later discussion, we explicitly describe
the structure of the interactions relevant for these processes in the 
following.

When the Higgsino mass $\mu$ has a nonvanishing phase, the elements of 
the neutralino mixing matrix $N$ have an imaginary part in general. 
The interactions of the neutral Higgses with the neutralinos $\chi^0_i$
($i$ $=$ 1, 2, 3, 4) 
in the presence of CP violation have the following structure 
\begin{eqnarray}
{\cal L}_{\chi^0\chi^0 H^0} & = & 
\frac{1}{2}\sum_{r=1}^3 \sum_{i,j=1}^4 \overline{\chi}^0_i 
  \left(C_S^{\chi^0_i\chi^0_j H^0_r} - C_P^{\chi^0_i\chi^0_j H^0_r} 
\gamma_5\right) \chi^0_j H^0_r. 
\label{eqn:int_h0_nn}
\end{eqnarray}
In the absence of CP violation, either the scalar coupling $C_S$ or the 
pseudoscalar coupling $C_P$ in eq.~(\ref{eqn:int_h0_nn}) 
is vanishing for every mass eigenstate $H^0_r$ \cite{MSSM}. 
Namely, the scalar coupling $C_S$ is vanishing for $H^0_3$, and
the pseudoscalar coupling $C_P$ is vanishing for $H^0_1$ and $H^0_2$. 
In the presence of CP violation, however, 
both $C_S$ and $C_P$ are nonzero for every mass eigenstate. 
These couplings can be written as 
\begin{eqnarray}
C_S^{\chi^0_i\chi^0_j H^0_r} & = & {\rm Re}(G_r^{ij}), \nonumber \\
C_P^{\chi^0_i\chi^0_j H^0_r} & = & i \, {\rm Im}(G_r^{ij}), 
\label{eqn:c_nnh}
\end{eqnarray}
where 
\begin{eqnarray}
G_r^{ij} & = & \frac{1}{2}(g' N_{i1}-gN_{i2}) 
 [ N_{j3}O_{r1}-N_{j4}O_{r2}  \nonumber \\
  &  &  \hspace{2cm} 
        + \, i O_{r3}(N_{j3}\sin\beta - N_{j4}\cos\beta) ] 
  \, + \, (i \leftrightarrow j). 
\label{eqn:g_nnh}
\end{eqnarray}
Note that the couplings in eq.~(\ref{eqn:c_nnh}) are vanishing 
if the lightest neutralino $\chi$ ($=$ $\chi^0_1$) 
is a pure bino or a pure higgsino. 
However, such a case can not take place in practice since 
there always exists a finite mixing between the bino and the higgsino.

The $W$-boson--$W$-boson--neutral Higgs boson interaction is given by  
\begin{eqnarray}
{\cal L}_{WWH^0} 
& = & \sum_{r=1}^3 
C^{WWH^0_r} H^0_r W_{\mu}^+ W^{-\mu}. 
\label{eqn:int_h0_ww}
\end{eqnarray}
The expression for the coefficient is 
\begin{eqnarray}
C^{WWH^0_r} & = & 
gm_W \left[ O_{r1}\cos\beta + O_{r2}\sin\beta \right], 
\label{eqn:c_wwh}
\end{eqnarray}
where $m_W$ is a mass of the $W$-boson. 
Numerically, the magnitude of $C^{WWH^0_r}$ is the largest for 
the lightest Higgs $H^0_2$, and the others are much smaller as
we explicitly see in Section \ref{sec:numerical-results}.

The chargino--neutralino--$W$-boson interactions are given by 
\begin{eqnarray}
{\cal L}_{\chi\chi^{\pm} W^{\mp}} 
& = & \sum_{k=1}^2 \overline{\chi^-_k} \gamma^{\mu}
\left(C_L^{\chi^+_k\chi W^-}P_L + C_R^{\chi^+_k\chi W^-}P_R \right) 
\chi W_{\mu}^- + \ {\rm h.c.} \ ,
\label{eqn:int_cnw}
\end{eqnarray}
where $P_L$ $=$ $(1-\gamma_5)/2$, $P_R$ $=$ $(1+\gamma_5)/2$ and 
\begin{eqnarray}
C_L^{\chi^+_k\chi W^-} & = & 
 - g \left( \sqrt{\textstyle{\frac{1}{2}}}N_{13}^*U_{k2} + N_{12}^*U_{k1}
               \right), 
 \nonumber \\
C_R^{\chi^+_k\chi W^-} & = & 
 g \left( \sqrt{\textstyle{\frac{1}{2}}}N_{14}V_{k2}^* - N_{12}V_{k1}^* 
               \right). 
\label{eqn:c_cnw}
\end{eqnarray}
The $2\times 2$ unitary matrices $U$ and $V$ 
diagonalize the chargino mass matrix ${\cal M_{\chi^-}}$ as
$U^* {\cal M_{\chi^-}} V^{-1}$ $=$ diag($m_{\chi_1^-}$, $m_{\chi_2^-}$)
\cite{NRR2}. 
Note that these couplings are vanishing in a pure bino limit. 

Finally, trilinear couplings for the neutral Higgs fields are given by
\begin{eqnarray}
{\cal L}_{H^0 H^0 H^0} 
& = & \sum_{r,s,t=1}^3 
\frac{1}{3!}C^{H^0_r H^0_s H^0_t} H^0_r H^0_s H^0_t. 
\label{eqn:int_hhh}
\end{eqnarray}
The coupling constant $C^{H^0_r H^0_s H^0_t}$ is written as 
\begin{eqnarray}
C^{H^0_r H^0_s H^0_t} & = & 
\frac{gm_Z}{4\cos \theta_W}
\Bigg[  
(O_{r1}\cos\beta-O_{r2}\sin\beta)
(O_{s2}O_{t2}-O_{s1}O_{t1}+O_{s3}O_{t3}\cos 2\beta) \nonumber \\
 & & 
\hspace{2cm} + \, ({\rm the \ other \ permutations \ of} \ r,s,t) 
\Bigg], 
\end{eqnarray}
where $m_Z$ is the $Z$-boson mass, and $\theta_W$ denotes the Weinberg angle.

%
\section{Neutralino pair annihilations with CP violation}
\label{sec:cross-section}
%

Neutralino pair annihilations are of importance when we examine
the relic density of the neutralino and its indirect detection. 
The complete expressions for the total cross section of 
neutralino pair annihilations in the case of no CP violation 
are found in Ref.~\cite{NRR2}. 
We have extended the analysis of Ref.~\cite{NRR2} to incorporate 
the effects of CP violation,  
and derived the full expressions for the neutralino 
annihilation cross section at the tree level with CP-violating phases, 
taking into account the modified interactions including 
(\ref{eqn:int_h0_nn}), (\ref{eqn:int_h0_ww}) and (\ref{eqn:int_cnw}). 
Calculation of the total pair annihilation cross section $\sigma$ 
involves a number of final states: 
\begin{eqnarray}
\chi \chi & \longrightarrow & f \bar{f}, \  
H^0_r H^0_s (r,s=1,2,3), \ H^+H^-, \nonumber \\
     & & W^+W^-, \ ZZ, \ W^{\pm}H^{\mp}, \ ZH^0_r (r=1,2,3). 
\end{eqnarray}
The complete set of the analytic expressions including all the final
states are quite lengthy and complicated, so we do not list the full result. 
Instead, we present only some of the results for 
the $W$-boson pair production $\chi\chi$ $\to$ $W^+W^-$ and
the lightest Higgs boson pair production $\chi\chi$ $\to$ $H^0_2 H^0_2$ 
which are essential in later discussion. 
In our numerical analysis in section \ref{sec:numerical-results}, however, 
we include all the contributions to the neutralino annihilation cross 
section at the tree level. 
Note that we can assume nonrelativistic neutralinos in the analysis of
positron flux, while we have to keep the exact form to calculate 
the relic density.

\begin{figure}[p]
\hspace*{-2cm} 
\includegraphics[width=15cm]{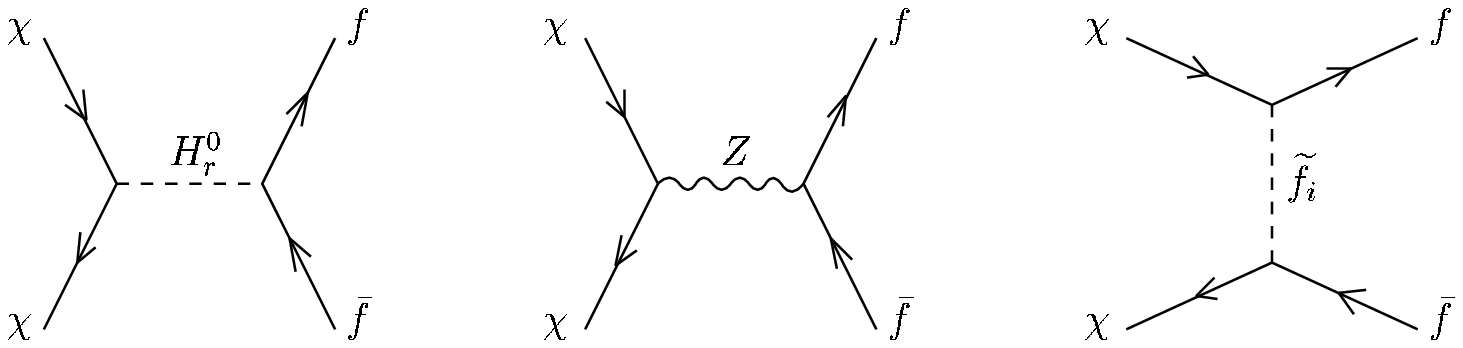}
\vspace{-20cm} 
\caption[cap:feyn-nn-ff]{
Feynman diagrams for $\chi\chi$ $\to$ $f\bar{f}$: 
the s-channel neutral Higgs boson exchange, the s-channel $Z$-boson exchange, 
and the t- and u-channel sfermion exchange. Note that the u-channel diagram
is not shown here. 
}
\label{fig:feyn-nn-ff}
\end{figure}

Before presenting the expressions with CP violation, 
let us briefly summarize some generic features on the cross sections 
in the case without CP violation. 
When the LSP is bino-like, 
the fermion pair production $\chi\chi \to f\bar{f}$ 
($f$ $=$ $u$, $c$, $t$, $\cdots$) typically gives a dominant contribution
to the total pair annihilation cross section for nonrelativistic 
neutralinos. 
There are three diagrams which contribute to this process: 
neutral Higgs boson exchange, $Z$-boson exchange and sfermion exchange
(see Fig.~\ref{fig:feyn-nn-ff}). 
The s-wave amplitude of this process for nonrelativistic neutralinos 
is always suppressed by a fermion mass (except for the top quark). 
Because of the s-wave suppression, the cross section of a light fermion
pair production from nonrelativistic neutralino annihilations is 
much smaller than that for a heavy fermion. 
If the neutralino is heavier than the $W$-boson and/or the lightest Higgs
boson, the processes $\chi\chi$ $\to$ $W^+W^-$ and/or 
$\chi\chi$ $\to$ $H^0_2 H^0_2$ open up. 
However, for a bino-like LSP, these processes give 
smaller contributions than that for the $f\bar{f}$ final state
in the case without CP violation.

On the other hand,  
when the LSP is higgsino-like, 
the $W$-boson pair production $\chi\chi$ $\to$ $W^+W^-$
gives a dominant contribution if the neutralino is heavy enough. 
This process involves three diagrams: 
neutral Higgs boson exchange, $Z$-boson exchange and chargino exchange
(see Fig.~\ref{fig:feyn-nn-ww}). 
Since this contribution has no s-wave suppression, the neutralino 
pair annihilation cross section is much enhanced so that the resultant 
relic density of the neutralino is typically too small to satisfy
the WMAP constraint (\ref{eqn:WMAP}).

However, in the presence of CP violation, new features appear in 
the $W$-boson pair production for a bino-like LSP. 
In order to see this fact, let us present essential parts of 
the analytic results for the cross section of $\chi\chi$ $\to$ $W^+W^-$ 
in the following.

The analytic expression for the neutral Higgs exchange contribution 
to $\chi\chi \to WW$ is given by 
\begin{eqnarray}
 \sigma_{WW}^{(H)} v & = &
\frac{\beta_{W}}{32\pi m_{\chi}^2}
   \frac{s^2-4m_W^2 s+12m_W^4}{8m_W^4}  \nonumber \\
 & & 
  \times 
  \left[ (s-4m_{\chi}^2) 
  \left| \sum_{r=1}^3 \frac{C^{WWH^0_r} C_S^{\chi\chi H^0_r}}
  {P_{H^0_r}(s)} \right|^2 
+ s \left|\sum_{r=1}^3 \frac{C^{WWH^0_r} C_P^{\chi\chi H^0_r}}
  {P_{H^0_r}(s)} \right|^2 \right], 
\label{eqn:sigmav_ww_h}
\end{eqnarray}
where 
\begin{eqnarray}
P_{H^0_r}(s) & = & s-m_{H^0_r}^2+i\, \Gamma_{H^0_r}\, m_{H^0_r}, 
\end{eqnarray}
$m_{\chi}$ is a mass of the lightest neutralino, 
$\Gamma_{H^0_r}$ denotes the decay width of $H^0_r$, 
$\beta_{W}$ $=$ $\sqrt{1-4m_W^2/s}$ 
is the velocity of the $W$-boson in the center of mass frame,
and $s$ is a Mandelstam variable. 
The relative velocity between the two colliding neutralinos is denoted 
by $v$. 
Numerically, the contribution from the lightest Higgs exchange ($r$ $=2$) 
is dominant as expected from eq.~(\ref{eqn:c_wwh}). 
Note that $C_P^{\chi\chi H^0_2}$ becomes nonzero in the presence of
CP violation. 
For nonrelativistic neutralinos $s$ $\to$ $4m_{\chi}^2$, the first term 
in the square bracket of eq.~(\ref{eqn:sigmav_ww_h}) vanishes. 
On the other hand, 
the second term includes the factor $s$ rather than $s-4m_{\chi}^2$. 
Therefore this term does not vanish in the nonrelativistic limit 
once CP violation is turned on. 


Similarly, the chargino exchange contribution and 
the interference term between the Higgs and the chargino 
exchange diagrams for $\chi\chi$ $\to$ $W^+W^-$ contain contributions
which do not vanish for $s$ $\to$ $4m_{\chi}^2$ in the presence of
CP violation. 
There are also other contributions to the $W$-boson pair production 
which include $Z$-boson exchange and possible interference terms, 
but we neglect them for the moment since they are typically subdominant. 
With all the relevant contibutions together, 
the expression for the total cross section of $\chi\chi$ $\to$ $W^+W^-$
for nonrelativistic neutralinos ($v$ $\to$ 0) is given by 
\begin{eqnarray}
\sigma_{WW} v \bigg|_{v \to 0}  & = &
\frac{\beta_{W}}{8\pi} 
\left[ 
(m_{\chi}^2-m_W^2)
\left| \sum_{k=1}^2 \frac{2C^W_{+k}}{\Delta^W_k} \right|^2
  \right. \nonumber \\
 & & 
\left. 
\hspace{1cm}
+ \, \frac{4m_{\chi}^4-4m_W^2m_{\chi}^2+3m_W^4}{2m_W^4}
\left|
\sum_{r=1}^3 \frac{C^{WWH^0_r} C_P^{\chi\chi H^0_r}}
  {P_{H^0_r}(4m_{\chi}^2)}
- \sum_{k=1}^2 \frac{2D^W_{-k}m_{\chi_k^-}}{\Delta^W_k}
\right|^2
\right], \nonumber \\
 & & 
\label{eqn:sigmav_ww_tot}
\end{eqnarray}
where $\Delta^W_k$ $=$ $m_W^2-m_{\chi}^2-m_{\chi_k^-}^2$ and 
\begin{eqnarray}
C^W_{+k} & = & 
\frac{1}{2}\left[ \bigg|C_L^{\chi^+_k\chi W^-}\bigg|^2 
                + \bigg|C_R^{\chi^+_k\chi W^-}\bigg|^2\right], 
  \nonumber \\
D^W_{-k} & = & 
i \, {\rm Im} \left[ 
C_R^{\chi^+_k\chi W^-}\bigg(C_L^{\chi^+_k\chi W^-}\bigg)^* \right]. 
\label{eqn:c-and-d_cnw}
\end{eqnarray}
In eq.~(\ref{eqn:sigmav_ww_tot}), the second term in the square bracket
represents the effect of CP violation, while the first term does not
vanish even without CP violation. 
Note that we include $Z$-boson exchange and possible interference terms
in the numerical analysis 
even though we neglected them in eq.~(\ref{eqn:sigmav_ww_tot}).

\begin{figure}[p]
\hspace*{-2cm} 
\includegraphics[width=10cm]{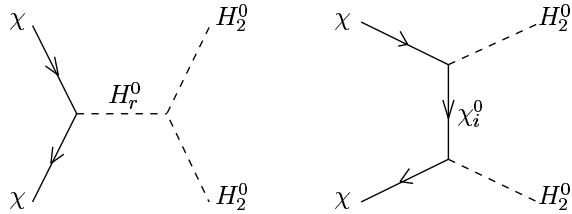}
\vspace{-19.5cm} 
\caption[cap:feyn-nn-hh]{
Feynman diagrams for $\chi\chi$ $\to$ $H^0_2 H^0_2$: 
the s-channel neutral Higgs boson exchange and 
the t- and u-channel neutralino exchange. 
}
\label{fig:feyn-nn-hh}
\end{figure}

Likewise, the lightest Higgs boson pair production 
$\chi\chi$ $\to$ $H^0_2 H^0_2$ 
involves contributions which are drastically enhanced by CP violation. 
This process occurs via two diagrams: 
neutral Higgs boson exchange and neutralino exchange
(see Fig.~\ref{fig:feyn-nn-hh}). 
The expression for the total cross section of this process 
for nonrelativistic neutralinos ($v$ $\to$ 0) is given by 
\begin{eqnarray}
\sigma_{H^0_2 H^0_2} v \bigg|_{v \to 0}  & = &
\frac{\beta_{H^0_2}}{16\pi} 
\left|
\sum_{r=1}^3 \frac{C^{H^0_2 H^0_2 H^0_r} C_P^{\chi\chi H^0_r}}
  {P_{H^0_r}(4m_{\chi}^2)}
- \sum_{i=1}^4 \frac{2D^{H^0_2}_{-i}m_{\chi^0_i}}{\Delta^{H^0_2}_i}
\right|^2, 
\label{eqn:sigmav_hh_tot}
\end{eqnarray}
where $\Delta^{H^0_2}_i$ $=$ $m_{H^0_2}^2-m_{\chi}^2-m_{\chi^0_i}^2$,
$\beta_{H^0_2}$ denotes the velocity of $H^0_2$ in the center of mass frame, 
$m_{\chi^0_i}$ is the mass of the neutralino $\chi^0_i$, and 
\begin{eqnarray}
D^{H^0_2}_{-i} & = & 
2i \, {\rm Im} \left[ 
C_S^{\chi^0_i\chi H^0_2}\bigg(C_P^{\chi^0_i\chi H^0_2}\bigg)^* \right]. 
\label{eqn:c-and-d_nnh}
\end{eqnarray}
Note that the cross section (\ref{eqn:sigmav_hh_tot}) is vanishing 
in the absence of CP violation.


The above expressions (\ref{eqn:sigmav_ww_tot}) and 
(\ref{eqn:sigmav_hh_tot}) are valid only for the nonrelativistic 
limit $s$ $\to$ $4m_{\chi}^2$. 
In our numerical calculation, however, 
we use the exact expression for general $s$ 
instead of the above expressions.

%
%
\section{Relic density of the neutralino}
\label{sec:relic-density}
%
In this section, we briefly review the formalism to compute 
the neutralino relic density in the MSSM with supersymmetric 
CP violation \cite{Jungman-etal}. 
The time evolution of the neutralino number density $n_\chi$ 
in the expanding Universe is described by the Boltzmann equation
\begin{eqnarray}
\frac{d n_\chi}{dt} + 3 H n_\chi & = & 
- \langle\sigma v \rangle 
\left[ n_\chi^2 - (n_\chi^{\rm eq})^2 \right], 
\label{eqn:Boltzmann-eq}
\end{eqnarray}
where $H$ is the Hubble expansion rate,
$n_\chi^{\rm eq}$ is the number density which the neutralino would have 
in thermal equilibrium,
and $\sigma$ is the total cross section of the neutralino pair annihilation 
into ordinary particles. 
The quantity $\langle \sigma v \rangle$ 
represents a thermal average of $\sigma v$.

In the early Universe, the neutralino is assumed to be in thermal 
equilibrium where $n_\chi$ $=$ $n_\chi^{\rm eq}$. 
As the Universe expands, the neutralino annihilation process freezes
out, and after that the number of the neutralinos in a comoving volume
remains constant. 
Using an approximate solution to eq.~(\ref{eqn:Boltzmann-eq}), 
the relic energy density $\rho_\chi$ $=$ $m_\chi n_\chi$
at present is given by
\begin{eqnarray}
\rho_\chi & = & \sqrt{\frac{4\pi^3 g_{*}G_N}{45}}
\left(\frac{T_\chi}{T_\gamma}\right)^3 T_\gamma^3 
\frac{1}{ \int_0^{x_F}dx \langle\sigma v \rangle },
\label{eqn:relic-density}
\end{eqnarray}
where $x$ $=$ $T/m_\chi$ is a temperature of the neutralino normalized 
by its mass, 
$g_{*}$ ($\approx 81$) represents the effective
number of degrees of freedom at freeze-out, 
and $G_N$ denotes the Newton's constant. 
$T_\chi$ and $T_\gamma$ are the present temperatures 
of the neutralino and the photon, respectively. 
The suppression factor $(T_\chi/T_\gamma)^3$ $\approx$ $1/20$ follows 
from the entropy conservation in a comoving volume \cite{reheating_factor}. 
The value of $x$ at freeze-out, $x_F$, 
is obtained by solving the following equation iteratively:
\begin{eqnarray}
x_F^{-1} & = & \ln \left( \frac{m_\chi}{2 \pi^3} \sqrt{\frac{45}{2g_* G_N}}
\langle\sigma v \rangle_{x_F} x_F^{1/2} \right). 
\label{eqn:freeze-out-temperature}
\end{eqnarray}
Typically one finds $x_F$ $\approx$ $1/20$ which implies that
the neutralinos are nonrelativistic at freeze-out.

The thermal average in eq.~(\ref{eqn:Boltzmann-eq}) should be 
carefully treated for accurate calculation of the relic density. 
In literatures, expansion of the thermal average in powers of 
the temperature $\langle\sigma v\rangle$ 
$\approx$ $a$ $+$ $bx$ is widely used. 
However, it is known that the expansion breaks down 
when $\sigma$ varies rapidly with the energy of the neutralinos, 
hence, in gerenal, one has to use the exact expression 
for the thermal average \cite{Gondolo-Gelmini}
\begin{eqnarray}
\langle\sigma v \rangle & = & 
\frac{1}{8 m_\chi^4 T K_2^2(m_\chi/T)} 
\int_{4 m_\chi^2}^\infty ds \, \sigma(s) (s-4m_\chi^2)\sqrt{s}
K_1\left(\frac{\sqrt{s}}{T}\right),
\label{eqn:thermal-average}
\end{eqnarray}
where $K_i$ ($i$ $=$ 1, 2) are the modified Bessel functions. 
Note that we should not take the nonrelativistic limit $v$ $\to$ 0
to calculate the cross section $\sigma$ in the integrand. 
In our analysis, we perform a numerical evaluation 
of the exact thermal average (\ref{eqn:thermal-average}) 
to obtain the relic density accurately.

In the present analyses, we neglect coannihilation effects
\cite{Griest-Seckel,Mizuta-Yamaguchi,ino-coan,stau-coan,stop-coan,NRR3,coan-recent}, 
although they are crucial when the next-lightest superparticle 
is nearly degenerated with the LSP. 
The investigation of the effect of CP violation on the coannihilation
cross sections is left for future work.

%
\section{Positrons from neutralino annihilations in the galactic halo}
\label{sec:positron}
%

Cosmic ray observations provide interesting probes 
for indirect detection of CDM. 
Even though the neutralino pair annihilations described in
section \ref{sec:relic-density} have already frozen out at present, 
neutralinos gravitationally accumulated in the galactic halo still 
can pair-annihilate to produce, e.g., cosmic $\gamma$-rays, 
neutrinos, antiprotons and positrions \cite{Jungman-etal,dm-recent-review}. 
Among various such observations, 
the High-Energy Antimatter Telescope (HEAT) 
reported an excess of cosmic ray positrons \cite{HEAT94-95}.  
Previous analyses on the positron flux in the MSSM without CP violation 
have suggested that some enhancement mechanism 
such as nontrivial distribution of CDM and/or so-called boost factors 
is necessary in order to explain the excess by neutralino LSP annihilations
\cite{positron-fraction-KT,positron-fraction-RS,positron-fraction-misc}.

We describe our procedure to evaluate the positrion flux 
from neutralino pair annihilations in the galactic halo in the following. 
The positron flux we measure can be written as
\cite{positron-fraction-RS,positron-fraction-KT} 
\begin{eqnarray}
\frac{dF_{+}}{dE} 
& = & \frac{\rho_0^2}{m_{\chi}^2}\int d\epsilon \, 
G(E,\epsilon) \sum_i ( \sigma_i v ) f_i(\epsilon), 
\end{eqnarray}
where $v=10^{-3}$ is a typical neutralino velocity in the
galactic halo, and 
$\rho_0=0.43$ GeV/cm${}^3$ represents a local halo dark matter density. 
Positrons produced in the halo are decelerated or accelerated 
during propagation in the halo until they are detected. 
The energy of a positron at detection is denoted by $E$, 
while that at production is denoted by $\epsilon$. 
The function $G(E, \epsilon)$ is a Green's function which describes 
propagation of positrons in the galactic halo. 
In our calculation, we use the Green's function in 
Ref.~\cite{positron-fraction-KT} with a containment time $\tau=10^7$ yr. 
The cross section $\sigma_i$ represents that for the process 
$\chi\chi$ $\to$ $i$, where the symbol $i$ runs over every possible
final state: $i$ $=$ $f\bar{f}$, $W^+W^-$, etc. 
Computation of the positron flux requires 
the cross sections $\sigma_i$ only for $v$ $=$ $10^{-3}$. 
In our numerical calculation of $\sigma_i$ for $v$ $=$ $10^{-3}$,
we use $s$ $\approx$ $4 m_{\chi}^2 (1+v^2)$. 
%

The function $f_i(\epsilon)$ is the positron energy spectrum at production
which originates from the process $\chi\chi$ $\to$ $i$. 
There are many contributions to $f_i(\epsilon)$, 
and some of them including hadronization are not calculable. 
In order to perform a detailed analysis, 
one has to utilize a Monte Carlo simulation code such as PYTHIA \cite{pythia}. 
In the present analysis, however, 
we content ourselves with a rough estimation to calculate $f_i(\epsilon)$, 
employing the methods used in 
Refs.~\cite{positron-fraction-RS,positron-fraction-KT} as follows.

The most energetic positron comes from 
the direct production  $\chi\chi$ $\to$ $e^+e^-$
where the positron has the maximal energy $\epsilon$ $\approx$ $m_{\chi}$. 
However the cross section is extremely small due to 
the s-wave suppression by the electron mass so that this contribution
is invisible. 
On the other hand, 
heavy fermion productions $\chi\chi$ $\to$ $c\bar{c}$, $b\bar{b}$,
$\tau^+\tau^-$ and $t\bar{t}$, 
may produce enough positrons as decay products 
(e.g., $\bar{b}$ $\to$ $\bar{c}W^{+ *}$ followed by
       $W^{+ *}$ $\to$ $e^+\nu_e$),
where the typical positron energy is $\epsilon$ $\approx$ $m_{\chi}/3$. 
Also, hadronization of quarks results in a shower of charged pions, 
and the contribution of the positrons from charged pion decay 
($\pi^+$ $\to$ $\mu^+$ $\to$ $e^+$) are evaluated using the data
from $e^+e^-$ collider experiments as in Ref.~\cite{positron-fraction-RS}.

For a neutralino heavier than the $W$-boson, 
positrons with energy $\epsilon$ $\approx$ $m_{\chi}/2$ can be produced 
by the $W$-boson production $\chi\chi$ $\to$ $W^+W^-$ followed by 
the decay $W^+$ $\to$ $e^+\nu_e$. 
There is also continuum positron radiation from 
muons, $\tau$-leptons and heavy quarks produced in the $W$-boson decay 
(i.e., $W^+$ $\to$ $\mu^+$ $\to$ $e^+$, 
       $W^+$ $\to$ $\tau^+$ $\to$ $e^+$, etc.). 
The contribution of positrons from decay of charged pions 
produced in the $W$-boson pair production 
is estimated as in Ref.~\cite{positron-fraction-KT}.
Similarly, contributions from the $Z$-boson pair production 
$\chi\chi$ $\to$ $ZZ$ followed by the decays $Z$ $\to$ $e^+e^-$, etc. 
are also included. 
It is known that, when the LSP is higgsino-like, 
the gauge boson final states $W^+W^-$ and $ZZ$ can give rise to 
enough positron excess \cite{positron-fraction-KT}. 
For the process including $H^0_2$ in the final state, 
positrons from $H^0_2$ $\to$ $b\bar{b}$ are taken into account.

For the cosmic ray electron flux, we use 
$\frac{dF_{-}}{dE}$ $=$ 0.07 $(E/\gev)^{-3.3}$ 
${\rm cm^{-2} sr^{-1} GeV^{-1} sec^{-1}}$ \cite{bkg-electron}. 
The background flux of cosmic ray positrons are estimated with 
$\frac{dF_{+}/dE}{dF_{+}/dE \,+ \,dF_{-}/dE}$ 
$=$ 0.02 $+$ $0.10 (E/\gev)^{-0.5}$ \cite{bkg-positron}.

%
%
\section{Numerical results}
\label{sec:numerical-results}
%

In this section, we present our numerical results. 
In all the figures, we fix some of the parameters 
in eq.~(\ref{eqn:mssm-parameters}) as 
\begin{eqnarray}
m_0=|A|=1 \,\tev, \ 
m_A=500 \, \gev, \ \tan\beta=5, 
\label{eqn:parameter-fix}
\end{eqnarray}
and consider variations of the remaining parameters as follows
\begin{eqnarray}
\theta_{\mu}= 0 - \pi, \ \theta_{A}= 0 - \pi, \nonumber \\
M_2= 100 - 700 \,\gev, \ |\mu|= 100 - 700 \,\gev. 
\label{eqn:parameter-var}
\end{eqnarray}

In what follows, 
we shall show that the process $\chi\chi$ $\to$ $W^+W^-$ 
or $\chi\chi$ $\to$ $H^0_2 H^0_2$ can give a
significant contribution to $\sigma v$ in the presence of CP violation. 
For this purpose, let us first present the magnitude of the relevant 
couplings in this process.

\begin{figure}[p]
\hspace*{-2cm} 
\includegraphics[width=15cm]{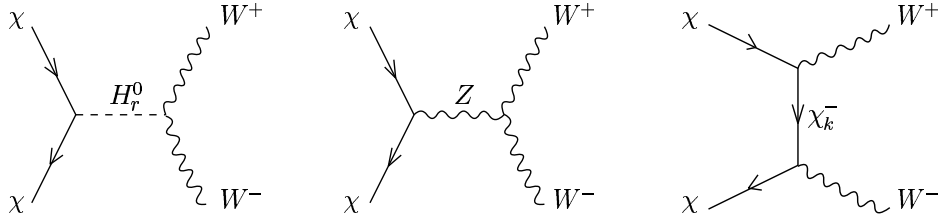}
\vspace{-20cm} 
\caption[cap:feyn-nn-ww]{
Feynman diagrams for $\chi\chi$ $\to$ $W^+W^-$: 
the s-channel neutral Higgs boson exchange, the s-channel $Z$-boson exchange,
and the t- and u-channel chargino exchange. 
}
\label{fig:feyn-nn-ww}
\end{figure}

\begin{figure}[p]
\includegraphics[width=15cm]{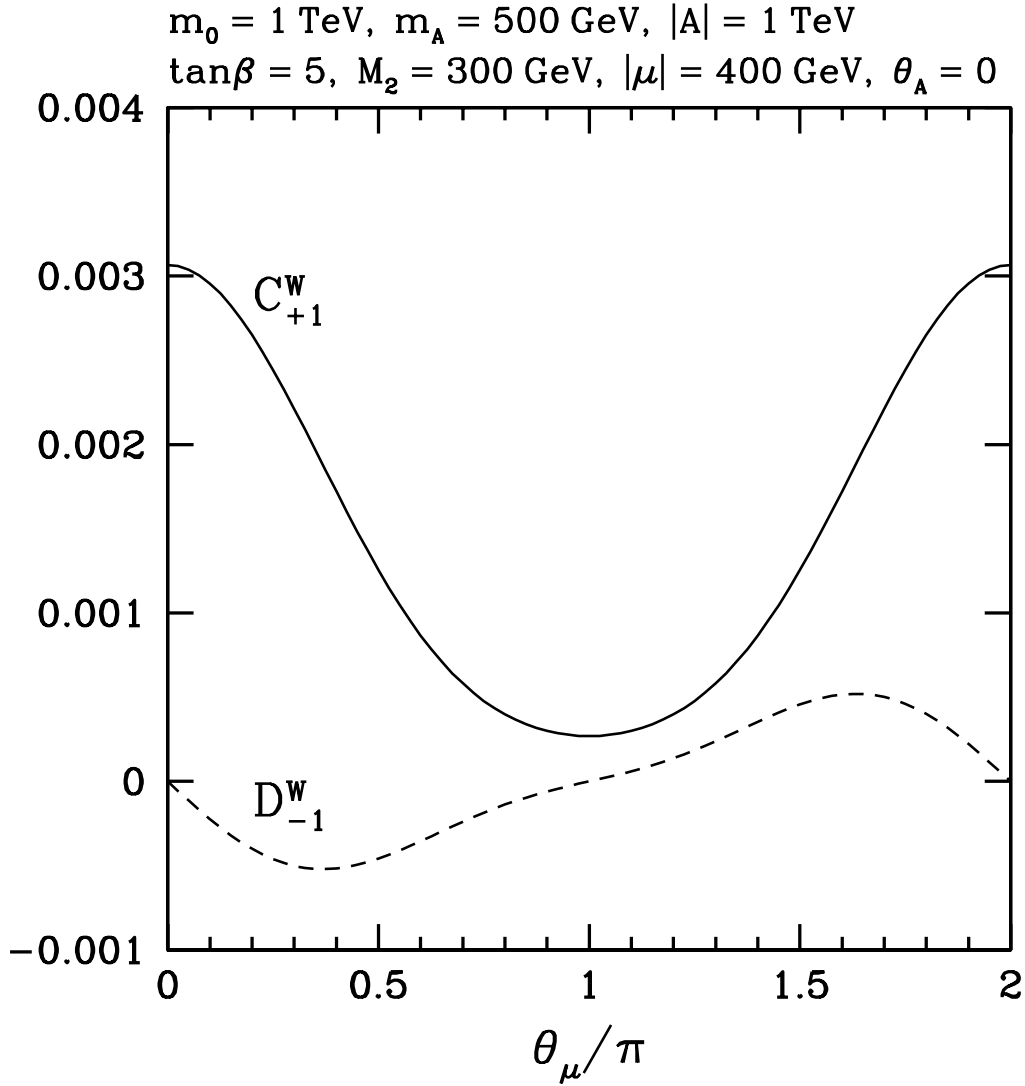}
\caption[cap:c_cnw]{
The variation of the quantities $C_{+1}^W$ and $D_{-1}^W$ defined 
in eq.~(\ref{eqn:c-and-d_cnw}) 
with the CP violating phase $\theta_{\mu}$. 
The relevant parameters are taken as $m_0$ $=$ $|A|$ $=$ 1 TeV, 
$m_A$ $=$ $500 \gev$, $\tan\beta$ $=$ 5, 
$M_2$ $=$ $300 \gev$, $|\mu|$ $=$ $400 \gev$ and $\theta_A$ $=$ 0. 
The dashed and solid curves correspond to the result for 
$C_{+1}^W$ and $D_{-1}^W$, respectively. 
In this case, the LSP is bino-like. 
}
\label{fig:c_cnw}
\end{figure}

We begin with a discussion of the t- and u-channel 
chargino exchange diagram (see Fig.~\ref{fig:feyn-nn-ww})
which is known to give a dominant contribution to $\chi\chi$ $\to$ $W^+W^-$
in the absence of CP violation. 
In Fig.~\ref{fig:c_cnw}, 
the quantities $C_{+1}^W$ and $D_{-1}^W$ defined 
in eq.~(\ref{eqn:c-and-d_cnw}) involved in the lighter
chargino exchange contribution to $\chi\chi$ $\to$ $W^+W^-$ are plotted 
as a function of the CP violating phase $\theta_{\mu}$ 
for $M_2$ $=$ $300 \gev$, $|\mu|$ $=$ $400 \gev$ and $\theta_A$ $=$ 0
with the other parameters fixed as in eq.~(\ref{eqn:parameter-fix}). 
The dashed and solid curves correspond to the result for 
$C_{+1}^W$ and $D_{-1}^W$, respectively. 
It is seen that both $C_{+1}^W$ and $D_{-1}^W$ have nontrivial dependence 
on $\theta_{\mu}$. 
The magnitudes of the coefficients in eq.~(\ref{eqn:chi-1}) 
for this choice of parameters 
are obtained as $|N_{11}|^2$ $\approx$ 0.97, $|N_{12}|^2$ $\lesssim$ 0.003,  
$|N_{13}|^2$ $\lesssim$ 0.02 and $|N_{14}|^2$ $\lesssim$ 0.006, 
hence the LSP is bino-like. 
Of course, the Higgs exchange diagram is absent if $\chi$ is a pure bino. 
However, $\chi$ actually has small but nonvanishing higgsino components 
due to bino--higgsino mixings even when $M_1$ is much smaller than $|\mu|$. 
The small higgsino components turn out to play a crucial role 
in the calculation of neutralino pair annihilation cross sections.

\begin{figure}[p]
\includegraphics[width=15cm]{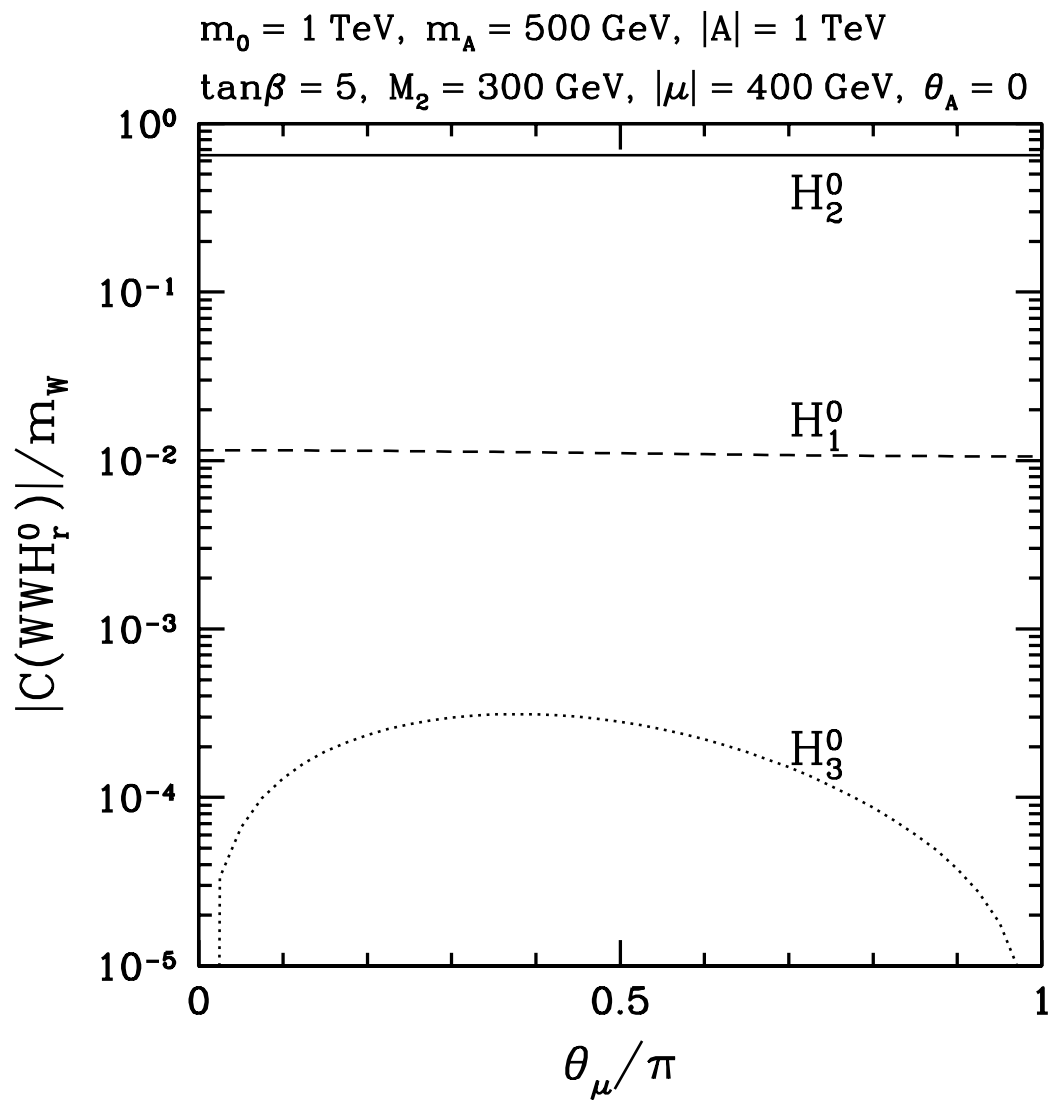}
\caption[cap:c_wwh]{
The absolute value of the coupling constants $C^{WWH^0_r}$ ($r$ $=$ 1, 2, 3) 
normalized by $m_W$ as a function of $\theta_{\mu}$ 
for the same choice of parameters as Fig.~\ref{fig:c_cnw}.
}
\label{fig:c_wwh}
\end{figure}

In the presence of CP violation, 
the largest contribution to $\chi\chi$ $\to$ $W^+W^-$ 
can be given by the neutral Higgs exchange diagram which includes 
three coupling constants $C^{WWH^0_r}$, $C_P^{\chi\chi H^0_r}$ and 
$C_S^{\chi\chi H^0_r}$ ($r$ $=$ 1, 2, 3). 
In Fig.~\ref{fig:c_wwh}, 
the dependence of the absolute value of the coupling constants $C^{WWH^0_r}$  
on $\theta_{\mu}$ is presented 
for the same choice of parameters as Fig.~\ref{fig:c_cnw}.
It is seen that the coupling  $C^{WWH^0_r}$ is the largest 
for the lightest Higgs $H^0_2$ independent of $\theta_{\mu}$.

\begin{figure}[p]
\includegraphics[width=15cm]{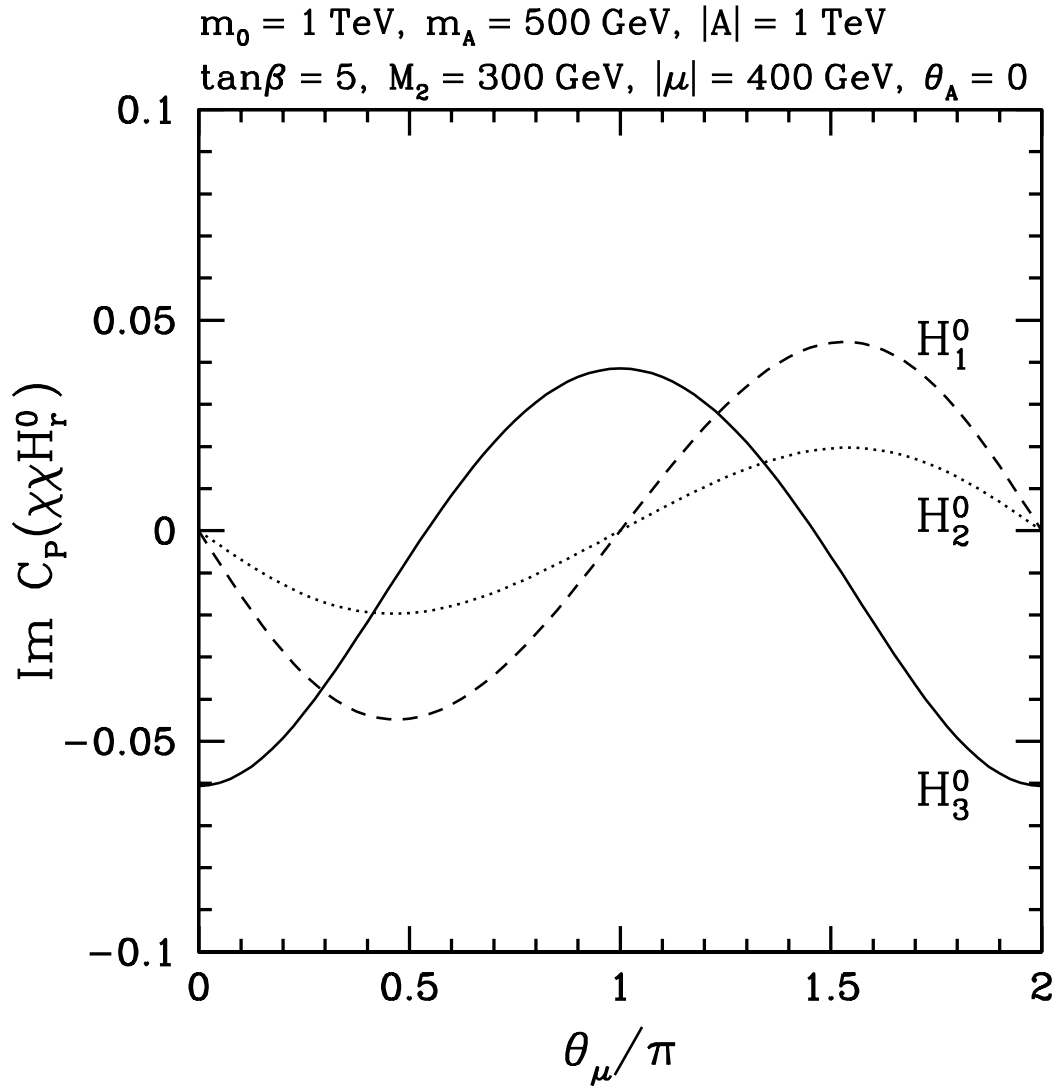}
\caption[cap:c_nnh_p]{
The imaginary part of the pure imaginary coupling constants 
$C_P^{\chi\chi H^0_r}$ ($r$ $=$ 1, 2, 3) 
as a function of $\theta_{\mu}$ for the same choice of parameters as 
Fig.~\ref{fig:c_cnw}.
}
\label{fig:c_nnh_p}
\end{figure}

On the other hand, the imaginary parts of 
the pure imaginary coupling constants $C_P^{\chi\chi H^0_r}$ 
($r$ $=$ 1, 2, 3) for the same choice of parameters are shown 
in Fig.~\ref{fig:c_nnh_p}. 
Without CP violation, only the coupling for $H^0_3$ is nonvanishing. 
In the presence of CP violation, however, the couplings for $H^0_1$ and
$H^0_2$ also become nonvanishing and even comparable to that for $H^0_3$. 
As these figures clearly show, the lightest Higgs exchange 
diagram gives the dominant contribution among the three Higgs exchange 
contributions. 
We do not show a result for the coupling $C_S^{\chi\chi H^0_r}$, 
since it is not essential for our discussion.

\begin{figure}[p]
\includegraphics[width=15cm]{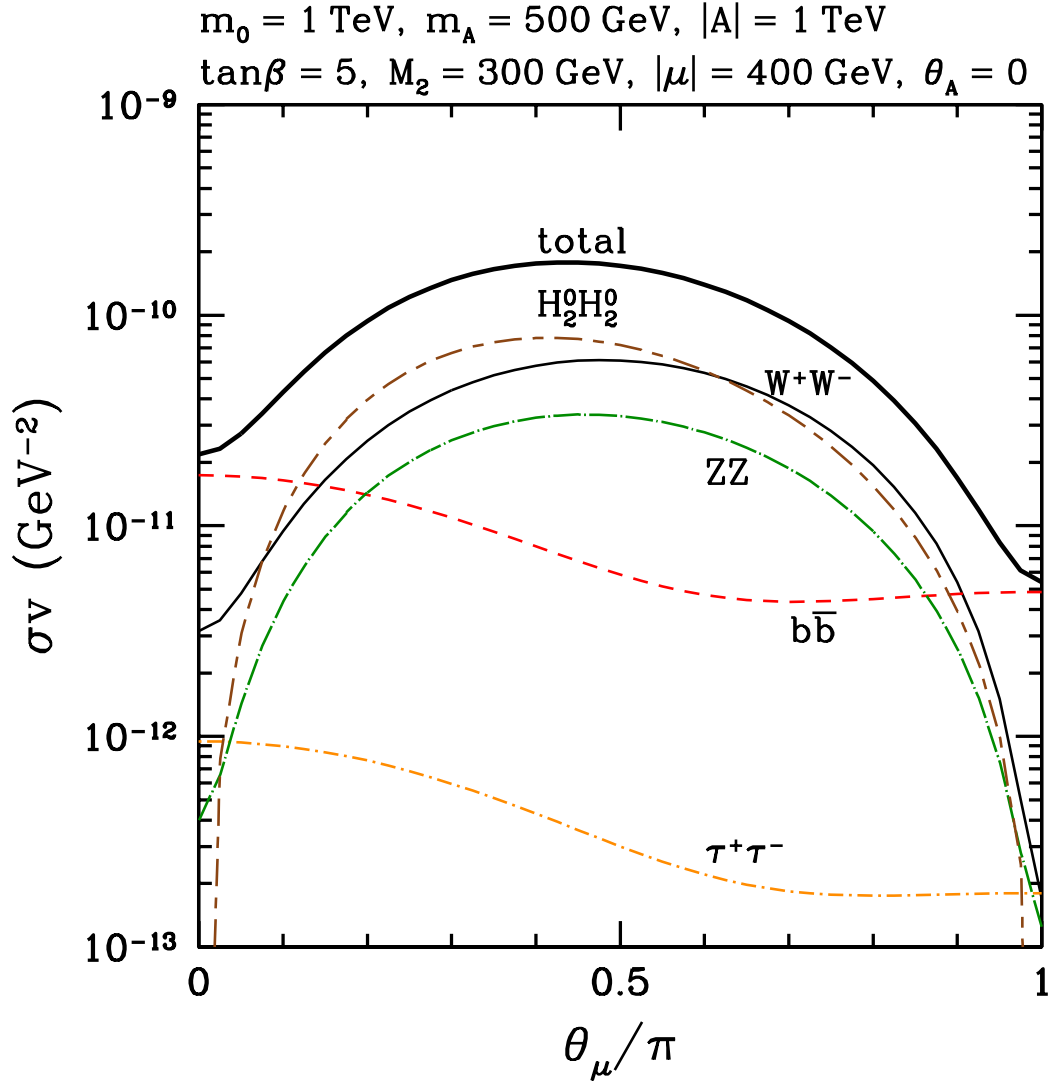}
\caption[cap:sigmav-thmu-bino]{
The cross section times relative velocity $\sigma v$ for $v=10^{-3}$ versus 
$\theta_{\mu}$ for the same choice of parameters as Fig.~\ref{fig:c_cnw}.
The solid, long dash-dot, short dashed, short dash-dot and 
short dash-long dash lines correspond to 
the contributions from 
$W^+W^-$, $ZZ$, $b\bar{b}$, $\tau^+\tau^-$ and $H^0_2 H^0_2$
final states, respectively. 
The bold solid line represents the sum of all the contributions. 
}
\label{fig:sigmav-thmu-bino}
\end{figure}

In Fig.~\ref{fig:sigmav-thmu-bino}, 
the cross section times relative velocity $\sigma v$ for $v=10^{-3}$  
is shown for the same choice of parameters as 
Fig.~\ref{fig:c_cnw}.
The solid, long dash-dot, short dashed, short dash-dot and 
short dash-long dash lines correspond to the contributions from 
$W^+W^-$, $ZZ$, $b\bar{b}$, $\tau^+\tau^-$ and $H^0_2 H^0_2$
final states, respectively. 
The bold solid line represents the sum of all the contributions. 
Without CP violation ($\theta_{\mu}$ $=$ 0, $\pi$), 
a fermionic final state, $b\bar{b}$, is dominant as usually expected 
for a bino-like LSP. 
In this case, the cross section of $\chi\chi$ $\to$ $W^+W^-$ is dominated
by the chargino exchange diagram in Fig.~\ref{fig:feyn-nn-ww}  
and the Higgs and $Z$-boson exchange contributions are negligible. 
The resultant contribution of the $W^+W^-$ final state is,
however, smaller than the $b\bar{b}$ contribution. 
For $\theta_{\mu}$ $\approx$ $\pi/2$, 
the lightest Higgs exchange contribution to the $W^+W^-$ production  
is much enhanced due to the effect of CP violation 
to dominate over the chargino exchange contibution. 
Because of this enhancement, the $W^+W^-$ contribution can be larger
than the $b\bar{b}$ contribution. 
Similar enhancement due to CP violation can be seen 
for $ZZ$ and $H^0_2 H^0_2$ final states. 
It is found that the $W^+W^-$ or $H^0_2 H^0_2$ final state 
gives the largest contribution for
$0.1\pi$ $\lesssim$ $\theta_{\mu}$ $\lesssim$ $0.9\pi$.

\begin{figure}[p]
\includegraphics[width=15cm]{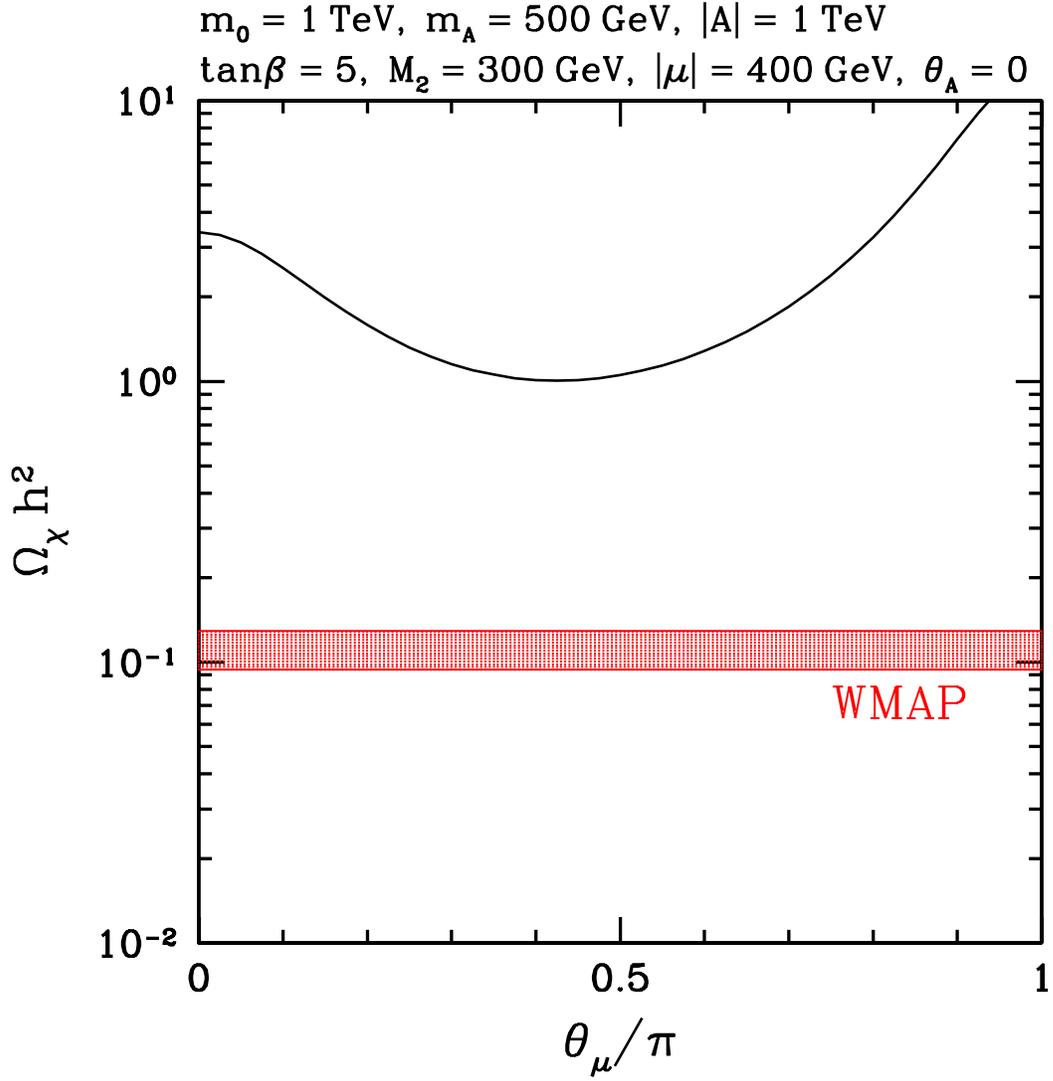}
\caption[cap:oh2-thmu-bino]{
The relic density $\Omega_{\chi}h^2$ versus 
$\theta_{\mu}$ for the same choice of parameters as Fig.~\ref{fig:c_cnw}.
In the shaded region, the relic density is consistent 
with the WMAP $2\sigma$ bound. 
}
\label{fig:oh2-thmu-bino}
\end{figure}

In Fig.~\ref{fig:oh2-thmu-bino}, 
the relic density $\Omega_{\chi}h^2$ is shown as a function of 
$\theta_{\mu}$ for the same choice of parameters as Fig.~\ref{fig:c_cnw}.
In the shaded region, the relic density is consistent with 
the $2\sigma$ allowed range of the WMAP result
\begin{eqnarray}
0.094 < \Omega_{\chi}h^2 < 0.129. 
\label{eqn:WMAP-2sigma}
\end{eqnarray}
It is found that the relic density is considerably suppressed for 
$\theta_{\mu}$ $\approx$ $\pi/2$. 
However, the relic density is still too large to satisfy the WMAP 
constraint even for $\theta_{\mu}$ $\approx$ $\pi/2$. 
In the presence of CP violation, the cross section is certainly much enhanced, 
but even this enhancement can not make bino-like dark matter consistent with
the cosmological constraint. 
In order to make the bino-like LSP cosmologically allowed, 
one has to resort to some other mechanism such as resonant annihilations
or coannihilations by tuning some parameter.

\begin{figure}[p]
\includegraphics[width=15cm]{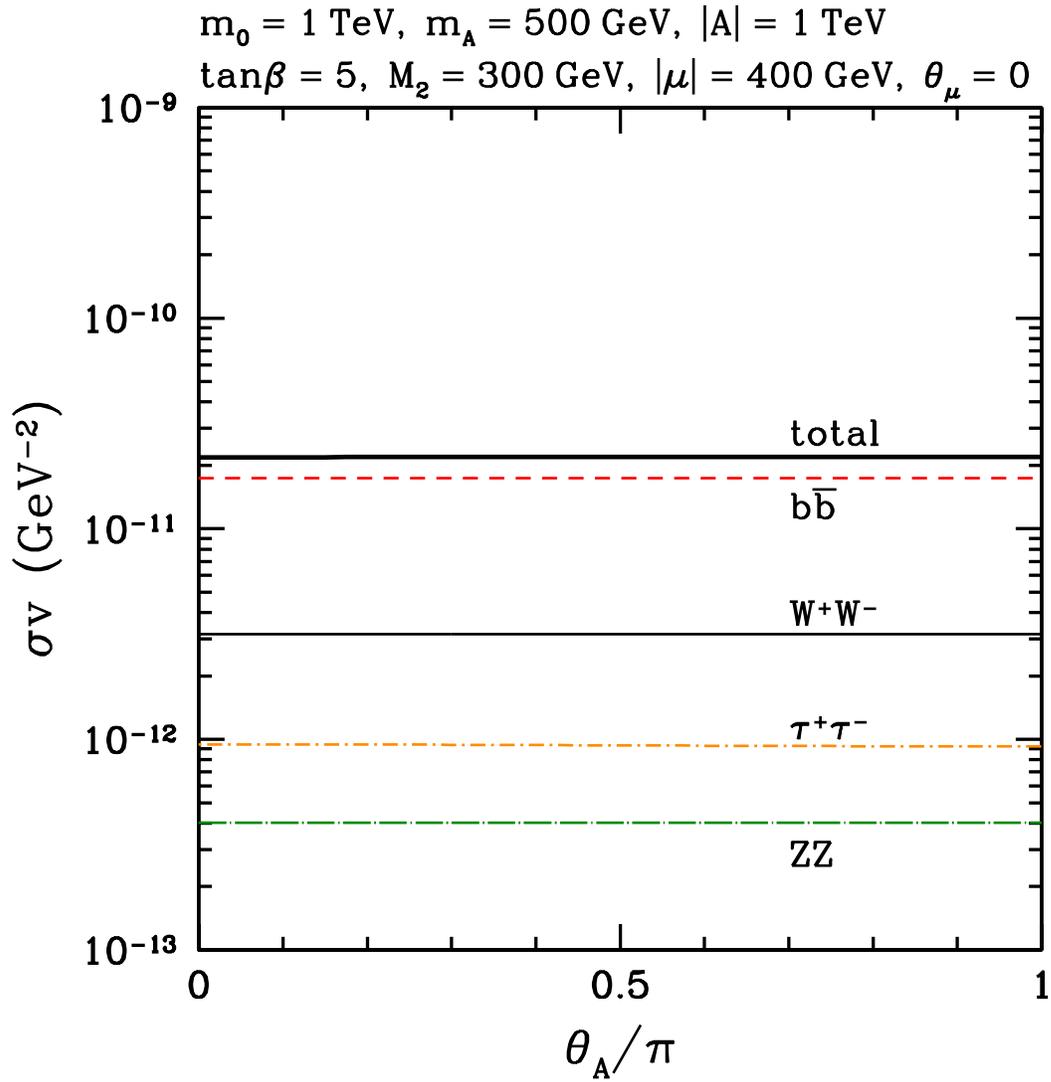}
\caption[cap:sigmav-tha0-bino]{
The cross section times relative velocity $\sigma v$ for $v=10^{-3}$ versus 
$\theta_{A}$ for the same choice of parameters as Fig.~\ref{fig:c_cnw}
but $\theta_{\mu}$ $=$ 0. 
}
\label{fig:sigmav-tha0-bino}
\end{figure}

The $\theta_{A}$ dependence of $\sigma v$ for $v=10^{-3}$ is shown in 
Fig.~\ref{fig:sigmav-tha0-bino} 
for the same choice of parameters as Fig.~\ref{fig:c_cnw}
but $\theta_{\mu}$ $=$ 0. 
It is seen that all the relevant contributions are insensitive to 
$\theta_A$. 
As a result, the relic density is also insensitive to $\theta_A$ 
($\Omega_{\chi}h^2$ $\approx$ 3.3). 
We have not found any significant dependence on $\theta_A$ even if
we vary $M_2$ and $\mu$ for our choice of the other parameters in 
eq.~(\ref{eqn:parameter-fix}).

\begin{figure}[p]
\includegraphics[width=15cm]{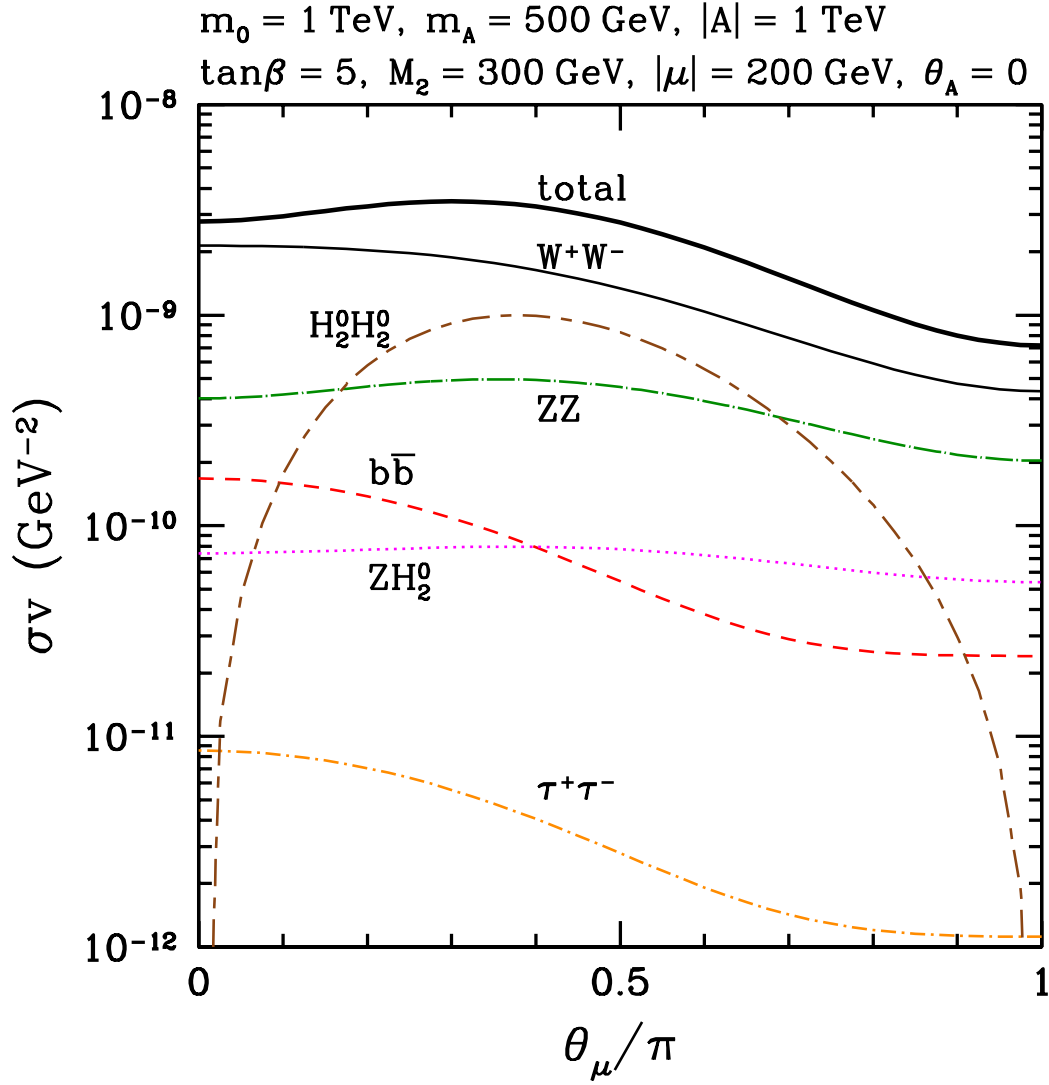}
\caption[cap:sigmav-thmu-mixed]{
The cross section times relative velocity $\sigma v$ for $v=10^{-3}$ versus  
$\theta_{\mu}$ for $M_2$ $=$ $300 \gev$ and $|\mu|$ $=$ $200 \gev$. 
The values of the other parameters are the same as Fig.~\ref{fig:c_cnw}.
The solid, long dash-dot, short dashed, long dashed, short dash-dot,
short dash-long dash and dotted lines correspond to 
the contributions from 
$W^+W^-$, $ZZ$, $b\bar{b}$, $t\bar{t}$, $\tau^+\tau^-$, $H^0_2 H^0_2$
and $Z H^0_2$ final states, respectively. 
The LSP in this case is a mixture of the bino and the higgsinos. 
}
\label{fig:sigmav-thmu-mixed}
\end{figure}

%
When the higgsino components are increased, the relic density shows 
different behavior. 
Fig.~\ref{fig:sigmav-thmu-mixed} represents 
the $\theta_{\mu}$ dependence of $\sigma v$ for $v=10^{-3}$ 
in the case of $M_2$ $=$ $300 \gev$ and $|\mu|$ $=$ $200 \gev$
where the LSP has sizable higgsino components. 
The values of the other parameters are the same as Fig.~\ref{fig:c_cnw}.
The magnitudes of the coefficients in eq.~(\ref{eqn:chi-1}) 
for this choice of parameters 
are obtained as $|N_{11}|^2$ $\approx$ 0.65, $|N_{12}|^2$ $\lesssim$ 0.03,  
$|N_{13}|^2$ $\approx$ 0.2 and $|N_{14}|^2$ $\approx$ 0.1, 
hence the LSP can be regarded as a mixture of the bino and the higgsinos. 
For such a mixed LSP, the magnitude of the couplings 
$C_L^{\chi^+_1\chi W^-}$ and $C_R^{\chi^+_1\chi W^-}$ in 
eq.~(\ref{eqn:c_cnw}) are increased 
compared with the case of a bino-like LSP. 
Because of this enhancement, the $W^+W^-$ final state gives a dominant 
contribution $\sigma v$ $\approx$ $10^{-9} \gev^{-2}$ for any $\theta_{\mu}$. 
The $ZZ$ and $H^0_2 H^0_2$ final states are also much enhanced and 
either of them gives the second largest contribution. 
However, the cross sections for the gauge boson final states $W^+W^-$ 
and $ZZ$ are less sensitive to $\theta_{\mu}$ for a mixed LSP than the 
case of a bino-like LSP, even though the $H^0_2 H^0_2$ final state
still has an enhancement for $\theta_{\mu}$ $\approx$ $\pi/2$. 
Therefore, the total cross section shows only 
mild dependence on $\theta_{\mu}$. 
One might expect that the $W^+W^-$ final state is
also enhanced as in the case for the $H^0_2 H^0_2$ final state. 
However, what actually happens is that there occurs 
a considerable cancellation between the Higgs exhange and
the chargino exchange diagrams for a mixed LSP
\footnote{In the first version of this paper, the sign of the interference 
term between the Higgs exhange and the chargino exchange diagrams 
in the process  $\chi\chi$ $\to$ $W^+W^-$ was wrong. 
Hence this cancellation was not seen in the first version.}.  
Because of this cancellation, the $W^+W^-$ final state 
shows only mild dependence on $\theta_{\mu}$. 
Similar cancellation occurs for the $ZZ$ final state as well.

\begin{figure}[p]
\includegraphics[width=15cm]{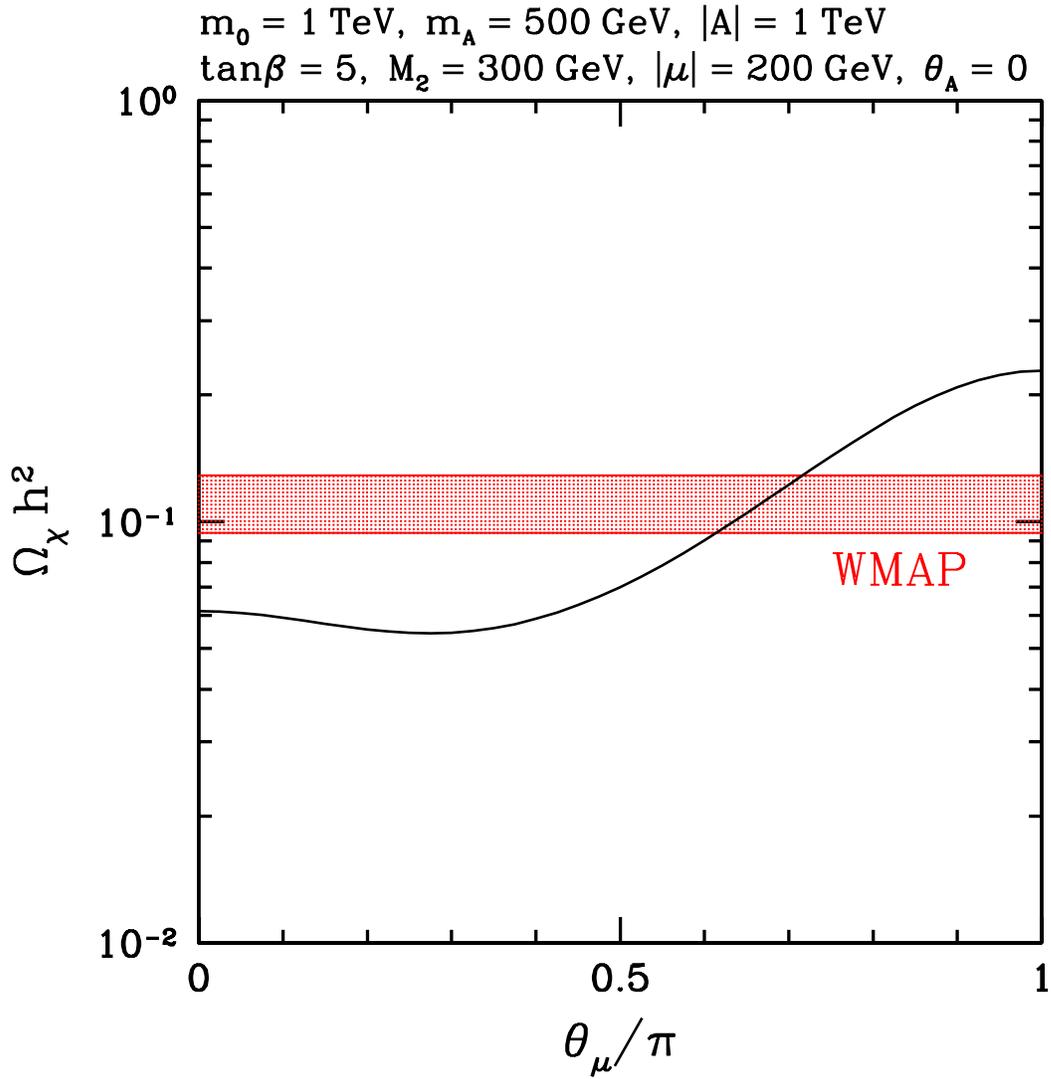}
\caption[cap:oh2-thmu-mixed]{
The relic density $\Omega_{\chi}h^2$ versus $\theta_{\mu}$ 
for the same choice of parameters as Fig.~\ref{fig:sigmav-thmu-mixed}.
}
\label{fig:oh2-thmu-mixed}
\end{figure}

The relic density for the same choice of parameters as
Fig.~\ref{fig:sigmav-thmu-mixed} is shown 
in Fig.~\ref{fig:oh2-thmu-mixed}. 
For a mixed LSP, 
the $\theta_{\mu}$ dependence of $\Omega_{\chi}h^2$ is much weaker
than the case of a bino-like LSP. 
However, it is found that the allowed region appear at 
$\theta_{\mu}$ $\approx$ $2\pi/3$. 
Thus CP violation can be essential to find cosmologically allowed regions
in the parameter space.

Let us briefly comment on a higgsino-like LSP. 
When $\chi$ is higgsino-like, 
the couplings $C_L^{\chi^+_1\chi W^-}$ and $C_R^{\chi^+_1\chi W^-}$ 
in eq.~(\ref{eqn:c_cnw}) have a maximal magnitude of order $g$ 
without any other small suppression factor,
and almost insensitive to $\theta_{\mu}$. 
In this case, the $W$-boson pair production gives the dominant contribution, 
and this leads to a much larger cross section $\sigma v$ $\approx$ 
$10^{-8} \gev^{-2}$ than that for a bino-like or a mixed LSP. 
Therefore the relic density for a higgsino-like LSP is too small to
satisfy the WMAP constraint, and almost insensitive to $\theta_{\mu}$.

\begin{figure}[p]
\includegraphics[width=15cm]{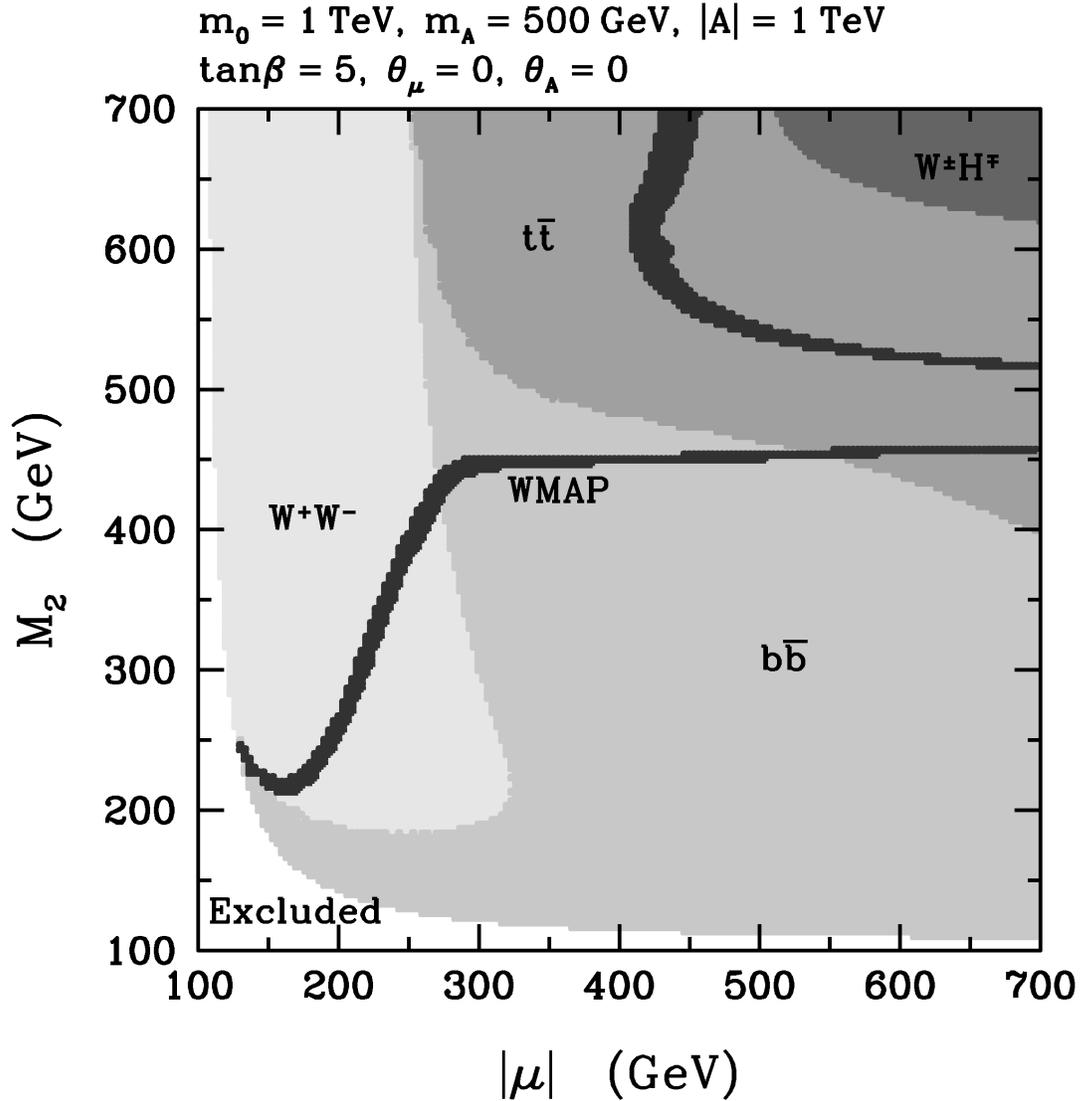}
\caption[cap:m2-mu-thmu0]{
The final states giving the largest contribution 
to $\sigma v$ for $v=10^{-3}$ in the ($|\mu|$, $M_2$) plane 
for $m_0$ $=$ $|A|$ $=$ 1 TeV, $m_A$ $=$ $500 \gev$, $\tan\beta$ $=$ 5,
$\theta_A$ $=$ 0 and $\theta_{\mu}$ $=$ 0. 
The regions where 
the final states $W^+W^-$, $b\bar{b}$, $t\bar{t}$ and $W^{\pm}H^{\mp}$ 
give the largest contribution are shown in different gray scales. 
From lighter to darker, each region corresponds to 
$W^+W^-$, $b\bar{b}$, $t\bar{t}$ and $W^{\pm}H^{\mp}$ 
in this order. 
In the darkest strip, the relic density is consistent with 
the WMAP $2\sigma$ result.  
The white region is excluded by the LEP limit 
on the chargino mass $m_{\chi^-_1}$ $>$ $104 \gev$ \cite{LEP-chargino} 
and the lightest Higgs mass $m_{H^0_2}$ $>$ $113 \gev$ \cite{LEP-higgs}. 
}
\label{fig:m2-mu-thmu0}
\end{figure}

Let us present a global map in the ($|\mu|$, $M_2$) plane. 
In Fig.~\ref{fig:m2-mu-thmu0}, 
the final states giving the largest contribution to $\sigma v$ for $v=10^{-3}$ 
are displayed in the ($|\mu|$, $M_2$) plane 
for $m_0$ $=$ $|A|$ $=$ 1 TeV, $m_A$ $=$ $500 \gev$, $\tan\beta$ $=$ 5,
$\theta_A$ $=$ 0 and $\theta_{\mu}$ $=$ 0. 
The regions where 
the final states $W^+W^-$, $b\bar{b}$, $t\bar{t}$ and $W^{\pm}H^{\mp}$
give the largest contribution are shown in different gray scales. 
The white region is excluded by the LEP limit 
on the chargino mass $m_{\chi^-_1}$ $>$ $104 \gev$ \cite{LEP-chargino} 
and the lightest Higgs mass $m_{H^0_2}$ $>$ $113 \gev$ \cite{LEP-higgs}. 
For a bino-like LSP ($M_2$ $\ll$ $|\mu|$), 
the $b\bar{b}$ final state typically gives 
the largest contribution, while for a higgsino-like LSP ($M_2$ $\gg$ $|\mu|$), 
the $W^+W^-$ final state is the largest. 
When the LSP is relatively heavy, the $t\bar{t}$ final state 
can be the largest. 
When both $M_2$ and $|\mu|$ are large ($\approx$ 700 GeV), 
the $W^{\pm}H^{\mp}$ final state is the largest. 
In the darkest strip in Fig.~\ref{fig:m2-mu-thmu0}, 
the relic density is consistent with the WMAP $2\sigma$ constraint. 
In the region $M_2$ $\approx$ $500 \gev$, 
where $2m_{\chi}$ $\approx$ $m_{H^0_3}$, 
there occurs resonant annihilation to the $b\bar{b}$ and $t\bar{t}$ 
final states via heavy Higgs boson exchange, 
so that the WMAP allowed region is extended to a large $|\mu|$ region 
for $M_2$ $\approx$ $500 \gev$. 
It is seen that aside from the resonant annihilation region, 
the cosmological constraint (\ref{eqn:WMAP-2sigma})
is satisfied for a mixed LSP. For a bino-like LSP, the relic density is
too large to satisfy the cosmological constraint (\ref{eqn:WMAP-2sigma}),
while it is too small for a higgsino-like LSP.

\begin{figure}[p]
\includegraphics[width=15cm]{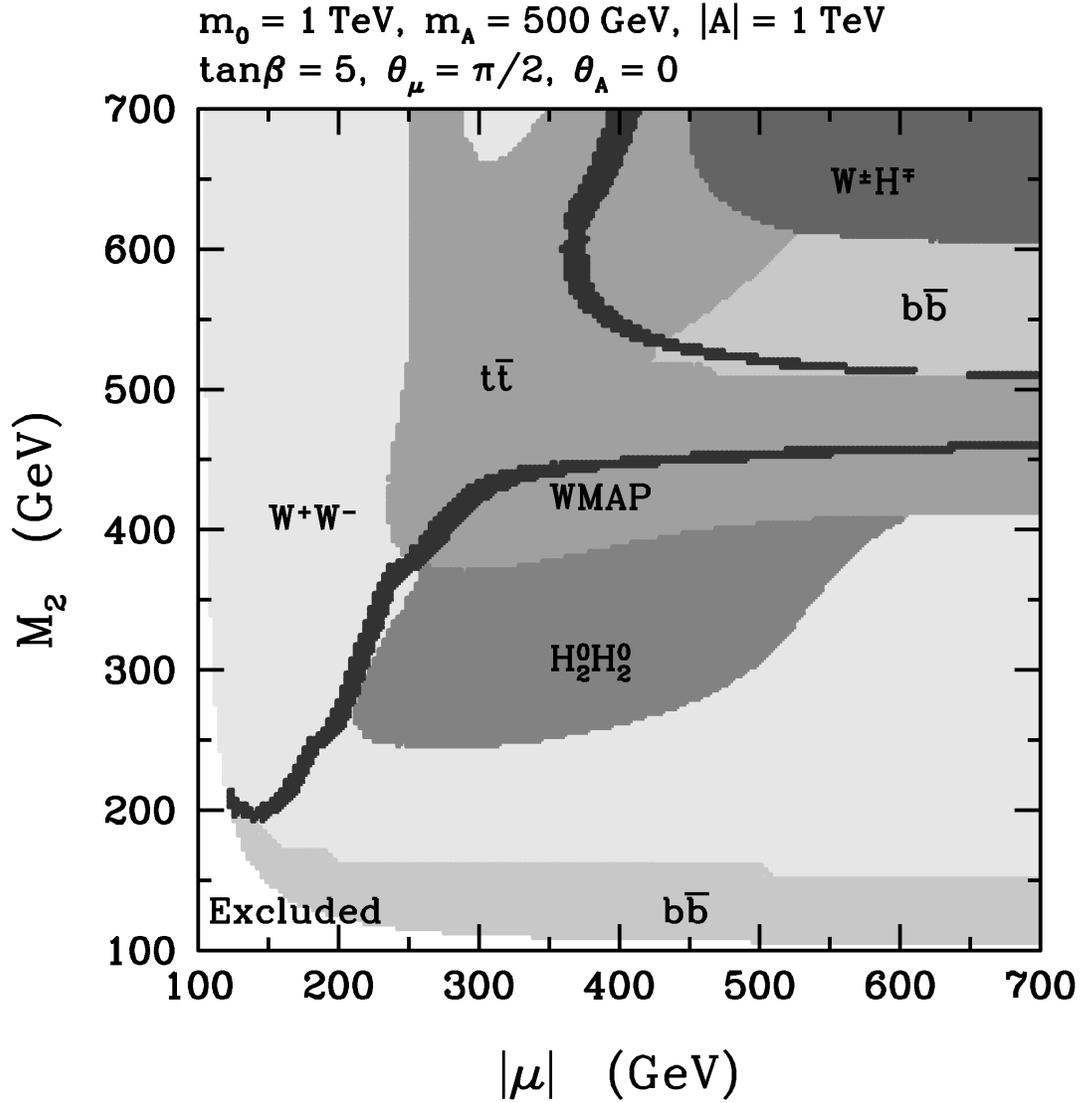}
\caption[cap:m2-mu-thmuPIov2]{
The same as Fig.~\ref{fig:m2-mu-thmu0} but for $\theta_{\mu}$ $=$ $\pi/2$. 
The regions where 
the final states $W^+W^-$, $b\bar{b}$, $t\bar{t}$, 
$H^0_2 H^0_2$ and $W^{\pm}H^{\mp}$ 
give the largest contribution are shown in different gray scales. 
From lighter to darker, each region corresponds to 
$W^+W^-$, $b\bar{b}$, $t\bar{t}$, $H^0_2 H^0_2$ and $W^{\pm}H^{\mp}$ 
in this order. 
}
\label{fig:m2-mu-thmuPIov2}
\end{figure}

The similar result to Fig.~\ref{fig:m2-mu-thmu0} 
but for $\theta_{\mu}$ $=$ $\pi/2$ is shown in Fig.~\ref{fig:m2-mu-thmuPIov2}.
It is found that, 
unlike the case without CP violation, 
the $W^+W^-$ production is significantly enhanced and 
gives the largest contribution even for a bino-like LSP. 
Note, however, that even this enhancement of the $W^+W^-$ final state 
can not save the bino-like region,
as the WMAP allowed region does not appear for $M_2$ $<$ $|\mu|$
except for the resonant annihilation region in Fig.~\ref{fig:m2-mu-thmuPIov2}. 
Also, there appears a region where the $H^0_2 H^0_2$ final state gives 
the largest contribution.

\begin{figure}[p]
\includegraphics[width=15cm]{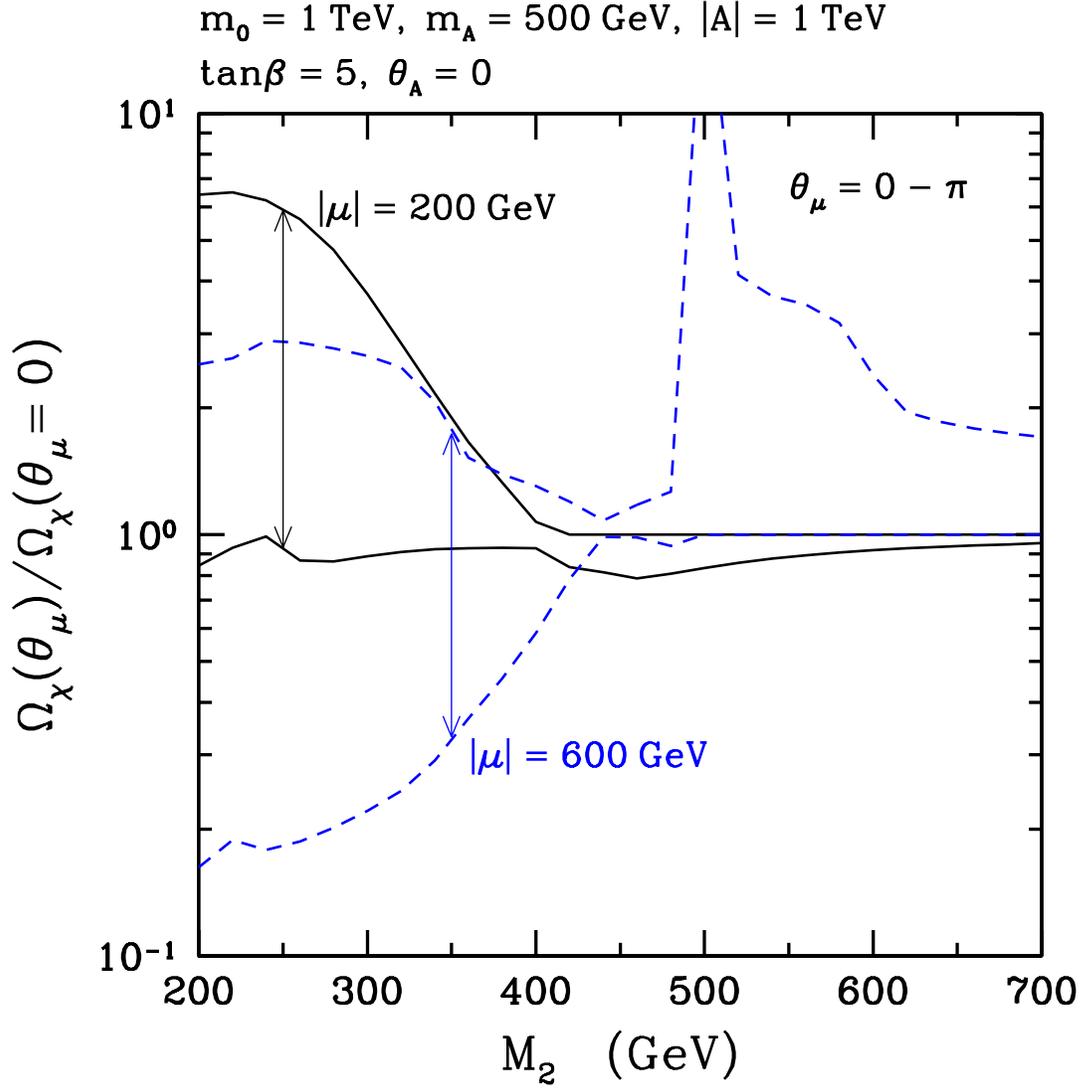}
\caption[cap:ratio-oh2]{
Variation of the relic density with $\theta_{\mu}$ 
normalized by that for $\theta_{\mu}$ $=$ 0, 
$\Omega_{\chi}(\theta_{\mu})/\Omega_{\chi}(\theta_{\mu}=0)$, 
as a function of $M_2$ for $m_0$ $=$ $|A|$ $=$ 1 TeV, 
$m_A$ $=$ $500 \gev$, $\tan\beta$ $=$ 5 and $\theta_A$ $=$ 0. 
Varing $\theta_{\mu}$ in the range 0 $<$ $\theta_{\mu}$ $<$ $\pi$, 
the relic density lies between the two solid lines for $|\mu|$ $=$ $200 \gev$. 
The region between the two dashed lines represents 
the corresponding result for $|\mu|$ $=$ $600 \gev$. 
}
\label{fig:ratio-oh2}
\end{figure}

In Fig.~\ref{fig:ratio-oh2}, 
variation of the relic density with $\theta_{\mu}$ 
normalized by that for $\theta_{\mu}$ $=$ 0, 
$\Omega_{\chi}(\theta_{\mu})/\Omega_{\chi}(\theta_{\mu}=0)$, 
is shown as a function of $M_2$ 
for $m_0$ $=$ $|A|$ $=$ 1 TeV, 
$m_A$ $=$ $500 \gev$, $\tan\beta$ $=$ 5 and $\theta_A$ $=$ 0. 
Varing $\theta_{\mu}$ in the range 0 $<$ $\theta_{\mu}$ $<$ $\pi$, 
the relic density lies between the two solid lines for $|\mu|$ $=$ $200 \gev$. 
The region between the two dashed lines represents 
the corresponding result for $|\mu|$ $=$ $600 \gev$. 
It is confirmed from this figure that the variation of the relic density 
with $\theta_{\mu}$ is typically small for a higgsino-like LSP.

\begin{figure}[p]
\includegraphics[width=15cm]{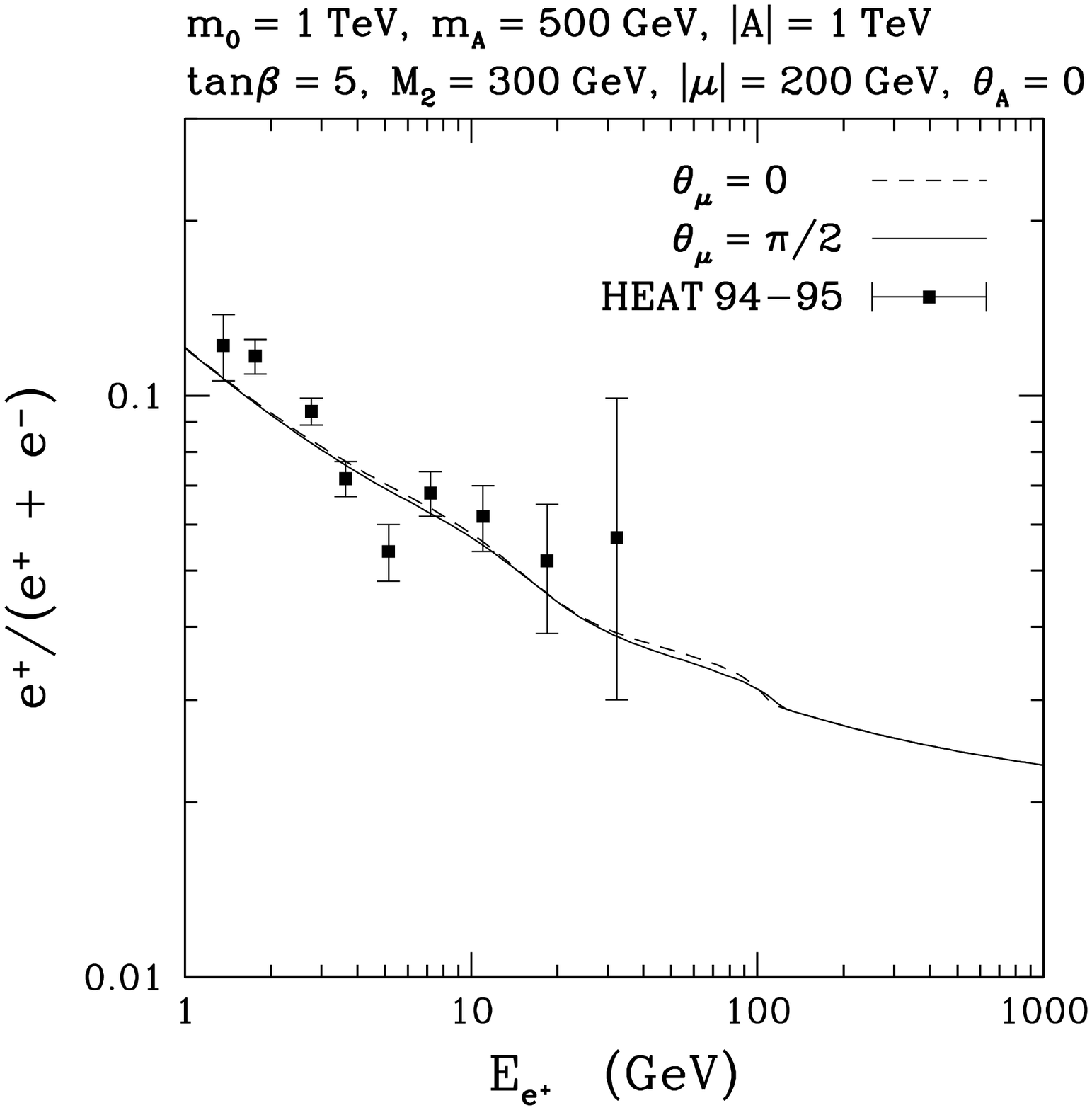}
\caption[cap:positron-flux-mixed]{
Positron fraction $e^+/(e^+ + e^-)$ versus positron energy $E_{e^+}$ 
for the same choice of parameters as in Fig.~\ref{fig:sigmav-thmu-mixed}. 
The dashed and solid lines correspond to the results for 
$\theta_{\mu}$ $=$ 0 and $\theta_{\mu}$ $=$ $\pi/2$, respectively. 
The points with error bars represent the data from HEAT measurement 
\cite{HEAT94-95}. 
}
\label{fig:positron-flux-mixed}
\end{figure}

\begin{figure}[p]
\includegraphics[width=15cm]{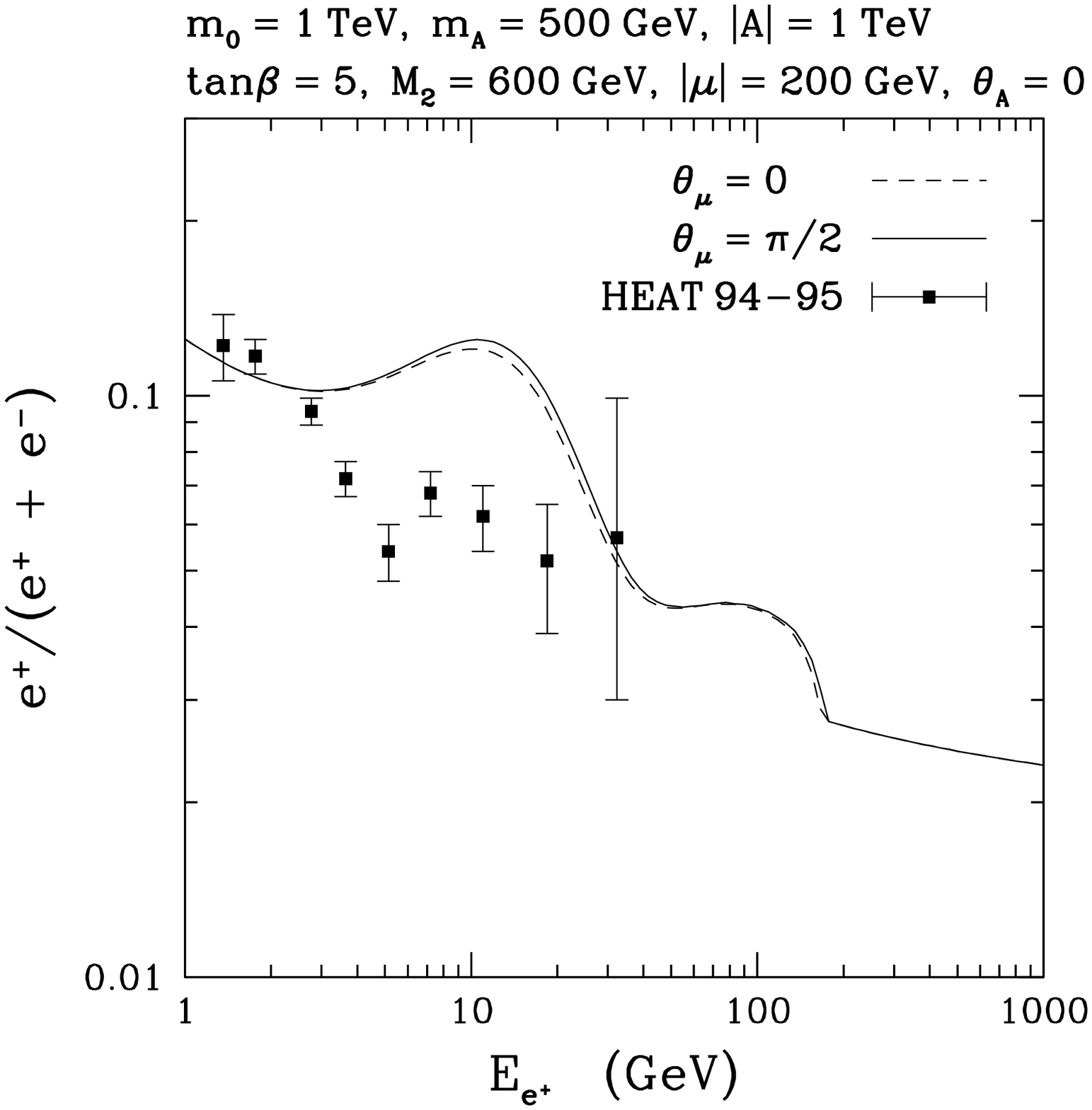}
\caption[cap:positron-flux-higgsino]{
The same as Fig.~\ref{fig:positron-flux-mixed} but 
for $M_2$ $=$ $600 \gev$. The LSP in this case is higgsino-like. 
}
\label{fig:positron-flux-higgsino}
\end{figure}

Finally we illustrate the effect of CP violation on the positron flux 
from neutralino annihilations in the galactic halo. 
The positron fraction $e^+/(e^+ + e^-)$ versus positron energy $E_{e^+}$ 
is shown in Fig.~\ref{fig:positron-flux-mixed}
for the same choice of parameters as in Fig.~\ref{fig:sigmav-thmu-mixed}. 
The LSP in this case is a mixture of the bino and the higgsinos.  
The dashed and solid lines correspond to the results for 
$\theta_{\mu}$ $=$ 0 and $\theta_{\mu}$ $=$ $\pi/2$, respectively. 
The points with error bars represent the data from HEAT measurement 
\cite{HEAT94-95}. 
For both $\theta_{\mu}$ $=$ 0 and $\theta_{\mu}$ $=$ $\pi/2$, 
the positron flux is dominated by the background 
so that the supersymmetric contribution from 
the LSP annihilations is almost invisible. 
The similar result for $M_2$ $=$ $600 \gev$ 
with the other parameters fixed as in Fig.~\ref{fig:positron-flux-mixed}. 
is shown in Fig.~\ref{fig:positron-flux-higgsino}. 
This choice of parameters corresponds to a higgsino-like LSP. 
In this case, 
the supersymmetric contribution gives a large excess 
due to enhancement of the cross sections for 
the $W^+W^-$ and $ZZ$ final states, as generically expected for 
a higgsino-like LSP \cite{positron-fraction-KT}. 
However, it is difficult to see the effect of CP violation, since 
the deviation of the solid line from the dashed one is very small
in Fig.~\ref{fig:positron-flux-higgsino}. 
In the case of a bino-like LSP as in Fig.~\ref{fig:c_cnw}, 
the neutralino pair annihilation cross section is smaller than that
for the case of a mixed LSP. Therefore the supersymmetric contribution 
to the positron flux is completely invisible.

In the present analysis, we have not included coannihilation effects 
\cite{Griest-Seckel,Mizuta-Yamaguchi,ino-coan,stau-coan,stop-coan,NRR3,coan-recent} 
in the calculation of the relic density. 
They are crucial in the case that the next-lightest superparticle 
is nearly degenerated with the LSP. 
Since this situation happens when the LSP is higgsino-like, 
our results on the relic density for a higgsino-like LSP 
may be altered if coannihilations are taken into account. 
The investigation of the effect of CP violation on the coannihilations
is left for future work.

%
%
\section{Conclusions}
\label{sec:conclusions}
%
We have studied pair annihilations of the neutralino dark matter 
in the minimal supersymmetric standard model with CP violation. 
We have considered the case that the higgsino mass and the trilinear scalar
couplings have CP-violating phases of order unity, 
taking a scenario that the scalar fermions in the first two generations 
are much heavier than those in the third generation to avoid a severe
constraint from experimental limits on electric dipole moments. 
It has been found that, when the lightest neutralino is bino-like, 
the cross section of the process $\chi\chi$ $\to$ $W^+W^-$ 
and $\chi\chi$ $\to$ $H^0_2 H^0_2$ 
for nonrelativistic neutralinos can be significantly enhanced 
by $\theta_{\mu}$, the phase of the higgsino mass. 
It follows that 
the relic density of the neutralino can be considerably suppressed 
by this effect. 
However, even this enhancement is not enough to make bino-like 
dark matter consistent with the cosmological constraint 
(\ref{eqn:WMAP-2sigma}).  
We have also discussed the effect of CP violation on the positron flux 
from neutralino pair annihilations in the galactic halo. 
It has been shown that the effect of CP violation is difficult to see 
in the positron flux.

%
%
%
\section*{Acknowledgments}
The author was supported in part by the Grant-in-Aid for Scientific
Research (No.16740150) from the Ministry of Education, Culture, Sports,
Science and Technology of Japan.
The author thanks G. Belanger, S. Kraml and S. Pukhov for pointing out 
the possible cancellation in the $W$-boson pair production 
with CP violation. 
%
%
%
%

%

\newpage 

%
%
%
%

%
%
%

%
%
%
%
%
%
%
%
\end{document}